\documentclass{article}
\usepackage[latin9]{inputenc}
\usepackage{amsmath}
\usepackage{amssymb}
\usepackage{graphicx}
\makeatletter

\newcommand{\Tr}{\operatorname{Tr}}
\newcommand{\re}{\operatorname{Re}}
\newcommand{\muhat}{\hat{\mu}}
\newcommand{\bra}{\langle}
\newcommand{\ket}{\rangle}
\newcommand{\err}[2]{\mbox{$\stackrel{\scriptstyle +#1}{\scriptstyle -#2}$}}
\makeatother

\begin{document}


\title{Enumerating Gribov copies on the lattice}

\author{Ciaran Hughes${}^{a}$\footnote{Current address: Department of
 Applied Mathematics and Theoretical Physics, University of Cambridge,
 Cambridge, CB3 0WA, UK},
 Dhagash Mehta${}^{a,b}$, 
Jon-Ivar Skullerud${}^a$\\
\mbox{}\\\mbox{}\\
${}^a${\em Department of Mathematical Physics, National University of Ireland
Maynooth},\\{\em Maynooth, Co Kildare, Ireland}\\%
${}^b${\em Department of Physics, Syracuse University, Syracuse, NY 13244,
USA}}


\maketitle

\begin{abstract}
In the modern formulation of lattice gauge-fixing, the gauge fixing
condition is written in terms of the minima or stationary points (collectively
called solutions) of a gauge-fixing functional. Due to the non-linearity
of this functional, it usually has many solutions called Gribov copies.
The dependence of the number of Gribov copies, $n[U]$ on the different
gauge orbits plays an important role in constructing the Faddeev--Popov
procedure and hence in realising the BRST symmetry on the lattice.
Here, we initiate a study of counting $n[U]$ for different orbits using
three complimentary methods: 1. analytical results in lower dimensions,
and some lower bounds on $n[U]$ in higher dimensions, 2. the numerical
polynomial homotopy continuation method, which numerically finds \textit{all}
Gribov copies for a given orbit for small lattices, and 3. numerical
minimisation ({}``brute force''), which finds many distinct Gribov
copies, but not necessarily all. Because $n$ for the coset SU($N_{c}$)/U(1)
of an SU($N_{c}$) theory is orbit-independent, we concentrate on
the residual compact U(1) case in this article and establish that
$n$ is orbit-dependent for the minimal lattice Landau gauge and
orbit-independent
for the absolute lattice Landau gauge. We also observe that contrary
to a previous claim, $n$ is not exponentially suppressed for the
recently proposed stereographic lattice Landau gauge compared to the
naive gauge in more than one dimension. 
\end{abstract}

\section{Introduction}

\label{sec:intro}

Though quantum field theory with the perturbative approach has been
extremely successful, there are many important physical phenomena,
such as quark confinement and dynamical chiral symmetry breaking in
quantum chromodynamics (QCD), which need a non-perturbative treatment.
To understand such a non-perturbative phenomenon, a genuine non-perturbative
approach to QCD is essential. An immensely useful approach to study
such non-perturbative phenomena is lattice field theory~\cite{Rothe:2005nw,Smit:2002ug,Montvay:1994cy}.
In this approach, the Euclidean space-time is discretised, so the
four-dimensional space-time integral is replaced by a discrete sum
over all lattice points while derivatives are replaced by finite differences.
Then, Monte Carlo methods used in statistical mechanics can be applied
to calculate the expectation values of observables.

In the continuum, a promising approach to study non-perturbative phenomena
in QCD is to study truncated systems of Dyson-Schwinger equations
(DSEs)~\cite{Alkofer:2000wg}, which are the equations of motion
for QCD Green's functions. Since each gauge configuration comes with
infinitely many equivalent physical copies, the set of which is called
a gauge-orbit of the gauge configuration, the generating functional
and hence any relevant quantity requires gauge-fixing to remove redundant
degrees of gauge freedom.

The DSE approach has a clear advantage in the low momentum region
of QCD. However, lattice QCD provides an opportunity to do \emph{first
principles} calculations of non-perturbative quantities in QCD: the
approximations involved in lattice QCD can be systematically removed,
unlike the truncations of DSEs. Thus, lattice simulations can provide
an independent check on the results obtained in the DSE approach.
A gauge field theory put on the lattice is manifestly gauge invariant,
i.e., one does not need to fix a gauge on the lattice to calculate
gauge invariant observables. However, to compare with DSE results,
gauge fixing is necessary, and it is mainly for this reason that lattice
Landau gauge studies have gained a large amount of interest recently.

The standard approach of gauge-fixing in the perturbative limit is
the Fad\-de\-ev--Popov (FP) procedure~\cite{Faddeev:1967fc}. There, a
gauge-fixing device which is called the gauge-fixing partition function,
$Z_{GF}$, is formulated. With an ideal gauge-fixing condition, $Z_{GF}$
is equal to one. This unity is inserted in the measure of the generating
functional so that the redundant degrees of freedom are removed after
appropriate integration. The generalisation of the FP procedure is
the Becchi--Rouet--Stora--Tyutin (BRST) formulation~\cite{Becchi:1975nq}.
The assumption that the gauge-fixing condition is ideal, i.e., the
gauge-fixing condition has a unique solution, is crucial here. However,
V.~N.~Gribov, in 1978, found that in non-Abelian gauge theories
a generalised Landau gauge-fixing condition treated non-perturbatively
would have multiple solutions, called Gribov or Gribov--Singer copies~\cite{Gribov:1977wm,Singer:1978dk,Alkofer:2000wg},
the effects of which should be properly taken into account.

There is one more obstacle on the lattice. If a procedure analogous
to the FP procedure is carried out on the lattice, the lattice analogue
of $Z_{GF}$ turns out to be zero~\cite{Neuberger:1986vv,Neuberger:1986xz}
due to a perfect cancellation among Gribov copies%
\footnote{See Refs. \cite{Fujikawa:1979ay,Hirschfeld:1978yq,Sharpe:1984vi}
for earlier attempts. %
}. Thus, the expectation value of a gauge-fixed observable turns out
to be of the indeterminate form $0/0$. This problem is known as the
Neuberger $0/0$ problem. In other words, BRST formulations on the
lattice can not be constructed. This severely hampers any comparison
with continuum DSE studies \cite{vonSmekal:2007ns,vonSmekal:2008es,vonSmekal:2008ws}.

On the lattice, Landau gauge fixing is usually formulated as a functional
minimisation problem. That is, instead of solving the lattice counterpart
of the gauge-fixing conditions, one numerically minimises a gauge-fixing
functional, whose first derivative with respect to gauge transformation
is the lattice counterpart of the gauge-fixing condition. The space
of minima, called the first Gribov region, also has many Gribov copies
but there is no cancellation among these Gribov copies hence the Neuberger
$0/0$ problem is avoided. However, the number of minima may be different
for different gauge-orbits making the corresponding $Z_{GF}$ orbit-dependent.
In this case, inserting $Z_{GF}$ in the partition function becomes
quite cumbersome. It is this orbit-dependence or orbit-independence
of the $Z_{GF}$ in various versions of lattice Landau gauge which is
the main focus of this paper.

Gribov copies not only play an important role in constructing the
BRST symmetry on the lattice, but it is also extensively argued that
they influence the infrared behaviour of the gauge dependent propagators
of gauge theories (See, e.g., \cite{Aouane:2011fv,Bornyakov:2011fn,Cucchieri:1997dx},
and \cite{Holdom:2007gx,Holdom:2009ws,Ilderton:2007dh,Ilderton:2007qy}
for a continuum perspective).

This paper is organised as follows. In Section~\ref{sec:gfix_lattice},
we introduce gauge-fixing on the lattice, fix the notation and set
up the general problem. We also introduce several versions of lattice
Landau gauge which we will study in the remainder of the paper. In
Section~\ref{sec:U1formalism} we formulate the gauge fixing problem
for the compact U(1) case. We also summarise the previously known
analytical results in one dimension and work out some analytical results
in higher dimensions. In Section \ref{sec:NPHC}, we introduce a novel
numerical method which finds all the stationary points of a given
potential provided that the corresponding stationary equations are
polynomial-like and have only finitely many solutions. We show how
to translate the lattice Landau gauge-fixing equations for the compact
$U(1)$ case into a system of polynomial equations, and then find
all the Gribov copies for small but non-trivial lattices. Section~\ref{sec:numerical}
contains the results from our numerical minimisation of the gauge
fixing functional. Finally, in Section~\ref{sec:discuss}, we summarise
our main results and discuss the implications of our findings.

\section{Gauge-fixing on the lattice}

\label{sec:gfix_lattice}

A gauge field theory can be studied nonperturbatively by discretising
the Euclidean space-time and putting the gauge and matter fields of
the theory on a four-dimensional space-time grid~\cite{Rothe:2005nw}.
The gauge fields are usually defined through link variables $U_{i,\mu}\in G$
where the discrete variable $i$ denotes the site index, $\mu$ is a directional
index and $G$ is the corresponding group of the theory. The relation
of the gauge-fields on the lattice $U_{i,\mu}\equiv U_{\mu}(x)$ to
their continuum counterpart $A_{\mu}(x)$, is given by
\begin{equation}
U_{\mu}(x)=\mathcal{P}\exp\Big[i\int_{x}^{x+\muhat}A_{\mu}(\xi)d\xi\Big]\,,
\end{equation}
 where $\mathcal{P}$ denotes path ordering.

The lattice expectation value of an observable $O$ is
\begin{equation}
\langle O\rangle=\frac{\int\prod_{i,\mu}dU_{i,\mu}\exp[-S[U]]O[U_{i,\mu}]}{\int\prod_{i,\mu}dU_{i,\mu}\exp[-S[U]]}\,,\label{eq:exp_value_of_O_lattice}
\end{equation}
 where $S[U]$ is the lattice action.

The action $S[U]$ is invariant under $U_{i,\mu}\rightarrow U_{i,\mu}^{g}=g_{i}^{\dagger}U_{i,\mu}g_{i+\muhat}$,
where the gauge transformations $g_{i}\in G$ are defined at each
lattice site. In other words, a set of randomly chosen $\{U\}$ comes
with infinitely many physically equivalent configurations the set
of which is called a gauge orbit and represented by the set $\{U\}$.
This gauge invariance makes the integration in the continuum counterpart
of Eq.~\eqref{eq:exp_value_of_O_lattice} ill-defined. In the continuum,
to get rid of this ambiguity, one can choose to take exactly one
representative (or more than one but a finite number) of each gauge-orbit.
This is called gauge fixing.

Choosing the representatives of each gauge-orbit can in practice be
done by imposing a constraint of the form $f(U^{g})=0$ in Eq.~\eqref{eq:exp_value_of_O_lattice}.
One takes the integral over the gauge orbit, $U^{g}$,
\begin{equation}
n[U]\equiv\int dg\,\det M_{FP}(g)\,\delta(f(U^{g}))=:Z_{GF}.\label{eq:Zgf}
\end{equation}
 where $Z_{GF}$ is called the gauge fixing partition function, and
$\det\, M_{FP}$ is the determinant of the Jacobian of the gauge-fixing
condition, called the Faddeev--Popov (FP) determinant. Here, $n[U]$ may be a function
of the gauge orbit $\{U\}$, that is, $n[U]=n[U^{g}]$. For non-Abelian
theories treated nonperturbatively, the surface $f(U^{g})=0$ will
in general have more than one solution, called {\em Gribov copies}.
If $n[U]$ is non-vanishing for all orbits, then inserting Eq.(\ref{eq:Zgf})
in \eqref{eq:exp_value_of_O_lattice} gives, after factoring out the
gauge volume,
\begin{equation}
\langle O\rangle=\frac{\int\prod_{i,\mu}dU_{i,\mu}\frac{1}{n[U]}\det M_{FP}(g)\,\delta(f(U^{g}))\exp[-S[U]]O[U_{i,\mu}]}{\int\prod_{i,\mu}dU_{i,\mu}\frac{1}{n[U]}\det M_{FP}(g)\,\delta(f(U^{g}))\exp[-S[U]]}\,,\label{eq:gauge_fixed_lat_exp_value}
\end{equation}
 provided that the operator $O[U]$ is gauge invariant. However, this
expression can also be used for gauge dependent operators, in which
case it defines the gauge fixed expectation value.

The most popular gauge is the Landau gauge, which can be formulated
as solving the following conditions in the continuum,
\begin{equation}
\partial_{\mu}A_{\mu}=0\;,\label{eq:landau_gauge_continuum}
\end{equation}
 or as a functional minimisation problem~\cite{Dell'Antonio:1991xt}
of the Landau gauge fixing functional
\begin{equation}
F_A[g] = \parallel A^{g}\parallel^{2}
  =-\int d^{4}x\Tr((gA_{\mu}g^{\dagger})^{2})\,,
\label{eq:continuum_gf_functional}
\end{equation}
whose first derivatives with respect to gauge parameters $g$ are
shown to give Eq.~\eqref{eq:landau_gauge_continuum}. The Hessian
matrix $M_{FP}$ of this functional is the FP operator.

The functional minimisation approach carries over straightforwardly
to the lattice formulation. The standard choice of the lattice Landau
gauge-fixing functional, in the following called the naive or direct
lattice Landau gauge functional, to be minimised with respect to the
corresponding gauge transformations $g_{i}$, is
\begin{equation}
F_{U}(g)=\sum_{i,\mu}(1-\frac{1}{N_{c}}\re\Tr g_{i}^{\dagger}U_{i,\mu}g_{i+\muhat}),\label{eq:general_l_g_functional}
\end{equation}
 for SU($N_{c}$) groups. Taking $f_{i}(g)=\frac{\partial F_{U}(g)}{\partial g_{i}}=0$
for each lattice site $i$ gives the lattice divergence of the lattice
gauge fields and in the naive continuum limit recovers the Landau
gauge $\partial_{\mu}A_{\mu}=0$. The corresponding $M_{FP}$ is the
Hessian matrix of $F_{U}(g)$, with respect to the gauge transformations.

Neuberger showed~\cite{Neuberger:1986vv,Neuberger:1986xz} that when
all the stationary points, not only minima, of the naive functional
are taken into account, $Z_{GF}$ turns out to be zero and the expectation
value of a gauge-fixed variable then is $0/0$. Schaden \cite{Schaden:1998hz}
interpreted the problem in terms of Morse theory and showed that $Z_{GF}$
calculates the Euler character $\chi$ of the group manifold $G$
at each site of the lattice. That is, for a lattice with $N$ lattice sites,
\begin{equation}
Z_{GF}=\sum_{i}\mbox{sign}(\det\, M_{FP}(g))=(\chi(G))^{N},\label{eq:Z_GF_is_Euler_char}
\end{equation}
 where the sum runs over all the Gribov copies. The group
manifold $G$ for compact U(1) is a circle $S^1$ and for SU($N_c$) it is
$S^3\times S^5 \times \dots \times S^{2 N_c-1}$. This result is derived
from the Poincar\'e--Hopf theorem which asserts that $\chi(\mathbb{M})$
is equal to the sum of signs of Hessian determinants at all critical
points of a non-degenerate height function which is a compact, differentiable
and orientable function from the manifold $\mathbb{M}$ to $\mathbb{R}$.
In the case of lattice Landau gauge fixing we can immediately identify
the gauge-fixing functional (\ref{eq:general_l_g_functional}) as
a height function, Gribov copies as the critical points and $M_{FP}$
as the corresponding Hessian matrix.

For compact U(1), $G=S^{1}$. As Eq.~(\ref{eq:general_l_g_functional})
can be viewed as a height function from $S^{1}\times S^{1}\dots\times S^{1}$
to $\mathbb{R}$, and since $\chi(S^{1})=0$, we have $Z_{GF}=0$.
As we will explicitly see in section~\ref{sec:U1formalism}, antiperiodic
boundary conditions in this case fix the global gauge freedom, so
the corresponding manifold is $(S^{1})^{N}$. For periodic boundary
conditions the corresponding manifold is $(S^{1})^{N-1}$. Thus, for
any boundary conditions, we have $Z_{GF}=0$. In fact, for any SU($N_{c}$),
and hence for the standard model group, the group manifolds are odd-dimensional
spheres for which $\chi$ is zero.

Following this interpretation, for SU(2) gauge theory, Schaden proposed
to construct a BRST formulation only for the coset space SU(2)/U(1)
for which $\chi\neq0$ (the residual compact U(1) symmetry was left
unaddressed). This procedure can be generalised to fix the gauge of
an SU($N_{c}$) lattice gauge theory to the maximal Abelian subgroup
$($U(1)$)^{N_{c}-1}$, since $\chi($SU($N_{c}$)/U(1)$)\neq0$ as
well. This indicates that the Neuberger $0/0$ problem for an SU($N_{c}$)
lattice gauge theory actually lies in (U(1)$)^{N_{c}-1}$, and
hence can be evaded if the problem for compact U(1) is evaded. This is why we concentrate our study on the compact
$U(1)$ case from now on.

We should emphasise here that there are no Gribov copies in continuum
Quantum Electrodynamics (QED) which is a U(1) gauge theory. More specifically,
on the lattice, the compactness of the gauge group introduces Gribov
copies for the compact U(1) theory. Thus, Gribov copies on the lattice
are purely lattice artefacts and so is the Neuberger 0/0 problem.
As argued above, we are interested in studying
Gribov copies of compact U(1) theory because the Neuberger 0/0 problem
apparently 
lies in the residual compact U(1) subgroup. Having said that, we should
also note that compact QED on the lattice may serve as a prototype
for theories based on compact groups such as SU($N_{c}$), and
hence studying compact QED on the lattice is important in its own
right. Compact QED in four dimensions exhibits two phases: a Coulomb
phase, with a massless photon, and a confined phase, which, although
unphysical, shares many qualitative
features with QCD.

To avoid the Neuberger $0/0$ problem, one may modify the gauge fixing
condition in some way. However, any such modified lattice gauge-fixing
should satisfy the following conditions:
\begin{enumerate}
\item The corresponding $n[U]$ should be orbit-independent. If this is
the case, then $n[U]$ cancels out in Eq.~\eqref{eq:gauge_fixed_lat_exp_value},
otherwise it must be computed for each configuration, which is usually
not feasible as it involves finding all solutions to $f_{i}(U^{g})=0$,
for $i$ running over all the lattice sites.
\item It should be possible to efficiently implement the corresponding gauge-fixing
numerically.
\item The additional gauge-fixing terms should not destroy the theory, e.g.,
the gauge-fixed action should be renormalisable.
\end{enumerate}
Below we discuss a few alternative gauge fixing procedures:

\textbf{Minimal lattice Landau gauge:} Here, instead of taking all
the stationary points of the gauge-fixing functional, one only considers
the space of minima, called the first Gribov region. Since $M_{FP}$
is positive definite for the minima, by definition, there is no cancellation
among the signs of determinants of $M_{FP}$ (i.e., $n[U]$ is just
the total number of local and global minima), and hence no Neuberger
$0/0$ is present there. This is also advantageous numerically, since
finding minima is easier than finding general stationary points. It
has also been shown \cite{Zwanziger:1989mf} that this restriction
can be written in terms of a renormalisable action with auxiliary
fields (for a review of this approach, see
\cite{Vandersickel:2012tz}). However, a crucial point is whether the
corresponding $n[U]$
is orbit-independent. If it is orbit-dependent then each orbit comes
with a different $n[U]$ and the functional integral becomes a function
of $n[U]$ and cumbersome to deal with. In the one-dimensional case
for compact $\mbox{U}(1)$, it has already been shown that the corresponding
$n[U]$ is orbit-dependent \cite{Mehta:2009,Mehta:2010pe}. In the
present paper, one of our goals is to verify this in the two-dimensional
case.

\textbf{Absolute lattice Landau gauge:} In this gauge, one further
restricts the gauge-fields to the space of global minima, called the
fundamental modular region (FMR). Here, in addition to evading the
Neuberger $0/0$ problem by avoiding the cancellation among the determinants
of $M_{FP}$ (i.e., $n[U]$ is nothing but the number of global minima),
we also expect the corresponding $n[U]$ to be orbit-independent,
and equal to 1 in the general case. It is also anticipated that the
set of configurations with degenerate global minima is a set of measure
zero which forms the boundary of the FMR. In other words, there are
no Gribov copies inside the FMR \cite{Zwanziger:1993dh,vanBaal:1997gu}.
In the one-dimensional compact $\mbox{U}(1)$ case, this was also
verified to be true \cite{Mehta:2009,Mehta:2010pe}. In the current
paper, we want to study this issue in the two-dimensional case. Having
said that, we emphasise that finding the global minimum of such functions
(which corresponds to spin glass model Hamiltonians) is known to be
a very difficult task and in most cases it is an NP hard problem.
Thus, in realistic cases, we can not expect to find the global minimum
using conventional numerical minimisation methods, and the best
one may do is to generate a number of minima and choose the `best
minimum' among these as an approximation to the global minimum (see
eg. \cite{Cucchieri:1997dx,Sternbeck:2005tk}). In addition to this,
the absolute lattice Landau gauge can not be stated in terms of algebraic
conditions, making it difficult to impose using the standard FP
procedure. In this paper, we provide evidence that $n[U]$ for this
gauge is indeed orbit-independent.

\textbf{Stereographic lattice Landau gauge:} A modification of the
group manifold of compact U(1), i.e., a circle $S^{1}$, via stereographic
projection at each lattice site was proposed and studied in Refs.~\cite{vonSmekal:2007ns,vonSmekal:2008es,Mehta:2009}.
For such a stereographically projected manifold the corresponding
$\chi$ is non-zero and thus the Neuberger $0/0$ problem is completely
avoided. Applying the same technique to the maximal Abelian subgroup
(U(1)$)^{N_{c}-1}$, the generalisation to SU($N_{c}$) lattice gauge
theories is possible when the odd-dimensional spheres $S^{2k+1}$,
$k=1,\dots,N_{c}-1$, of its parameter space are stereographically
projected to the real projective space $\mathbb{R}P(2k)$. It was
also shown using topological arguments that the number of Gribov copies
is exponentially suppressed for the stereographic compared to the
naive gauge for compact U(1), and that the corresponding $n[U]$ is
orbit-independent, at least for the one-dimensional lattice case.
It can be shown that the corresponding FP operator in this case is
generically positive (semi-)definite (i.e., all stationary points
are minima in this case, and there are no saddle points nor maxima)
and hence $n[U]$ is nothing but the total number of local and global
minima. The stereographic lattice Landau gauge is a promising alternative
from the lattice BRST symmetry point of view since it fulfils all
the above mentioned practical requirements, except that the orbit-independence
is yet to be confirmed for lattices in more than one dimension.

Interestingly, in lattice formulations of supersymmetric Yang--Mills
theories, non-compact parameterisations (similar to the stereographic
projection) of the gauge fields are used \cite{Catterall:2009it}.
The non-compact parametrisation, unlike the compact (group based)
parametrisation, apparently evades the sign problem in the
lattice versions of these supersymmetric
theories \cite{Catterall:2011aa,Galvez:2012sv}.
Recently, a deeper and direct connection between the
sign problem in lattice supersymmetric theories and the Neuberger
0/0 problem has been established \cite{Mehta:2011ud}: essentially, the
complete action of $\mathcal{N}=2$ supersymmetric
Yang-Mills theories in two dimensions can be shown to be a gauge-fixing
action, analogous to the Faddeev--Popov procedure, with the symmetry
now being a topological 
gauge symmetry. Thus, the corresponding partition function is
nothing but $Z_{GF}$ for the corresponding theory and hence the Neuberger
0/0 problem follows for the compact parametrisation. For the non-compact
parametrisation, the Neuberger 0/0 problem can be avoided due to 
topological arguments.

\textbf{Other approaches and efforts:} An alternative to fixing the
gauge completely would be to average over Gribov copies, as proposed
in Refs.~\cite{Parrinello:1990pm,Zwanziger:1990tn,Fachin:1991pu}, with
a weight proportional to the $\exp(-\beta_{GF}F[g])$, where $F[g]$ is
the gauge fixing functional of \eqref{eq:continuum_gf_functional} or
\eqref{eq:general_l_g_functional}.  This reformulation evades the
Neuberger problem.  In the original formulation, the average was over
the whole gauge orbit, and the (absolute) Landau gauge would only be
reproduced in the limit $\beta\to\infty$.  It has recently been shown
\cite{Serreau:2012cg} that if the average is taken only over
configurations satisfying the Landau gauge condition, the resulting
action is renormalisable and possesses a BRST-like symmetry.  This
class of gauges has been studied numerically \cite{Henty:1996kv};
however, it is expensive as it amounts to performing a second Monte
Carlo integral over Gribov copies on top of the usual lattice Monte
Carlo sampling.

An orthogonal approach
was recently proposed \cite{Maas:2009se}, in which the Gribov copies
are treated as a residual, nonperturbative gauge degree of freedom
which (at least in the continuum, infinite volume limit) can be fixed
by imposing additional conditions on the correlation functions. For
example, it is suggested that one may define a family of gauges, so-called
Landau-$B$ gauges, by defining a target value $B$ for the renormalised
ghost propagator at a specific momentum and choosing the Gribov copy
that gives a result closest to this value. A particular choice is
the max-$B$ gauge which selects the Gribov copy with the largest
value for $B$. Whether or to what extent it is possible to impose
such a condition is at present not clear, and little or nothing is
as yet known about the relation between the $B$ parameter and the
Landau gauge fixing functional.

Another way was put forward in Refs.\cite{Kalloniatis:2005if,Ghiotti:2006pm,vonSmekal:2008en}
using the ghost/anti-ghost symmetric Curci-Ferrari gauges . There
the argument used was that the Neuberger 0/0 problem could be extended
to include such non-linear gauges with their extended double-BRST
symmetry despite their quartic ghost self-interactions, which allow
the introduction of a mass term for ghosts. Such a Curci-Ferrari mass
would break the nilpotency of the BRST/anti-BRST charges which is
known to result in a loss of unitarity; however, this mass also serves
to regulate the Neuberger zeroes in a lattice formulation and expectation
values of observables can then be meaningfully defined in the limit
$m\rightarrow0$ via l'Hospital's rule.

On the continuum side, an explicit counting of Gribov copies was
carried out in \cite{Holdom:2007gx,Holdom:2009ws}. There, the SU(2) case was
considered, but restricted to static spherically
symmetric configurations only (in which case Landau and Coulomb gauge
conditions are identical).

This is not intended to be an exhaustive review of previous attempts
to address this subject.
For a pedagogical and thorough review of previous work on this topic,
we refer to the recent review \cite{Maas:2011se}.

\section{Lattice Landau gauge for compact U(1)}

\label{sec:U1formalism}

Since the group manifold of compact U(1) is topologically a circle
$S^{1}$, we can write the link variables and gauge transformations
in terms of angles $\phi_{i,\mu},\theta_{i}\in(-\pi,\pi]\mod2\pi$,
as $U_{i,\mu}=e^{i\phi_{i,\mu}}$and $g_{i}=e^{i\theta_{i}}$, respectively.
Thus, the naive gauge fixing functional Eq.~(\ref{eq:general_l_g_functional})
is reduced to
\begin{align}
F_{\phi}(\theta)
 & =\frac{1}{V}\sum_{i,\mu}\big(1-\cos(\phi_{i,\mu}+\theta_{i+\muhat}-\theta_{i})\big)
 \equiv \frac{1}{V}\sum_{i,\mu}(1-\cos\phi_{i,\mu}^{\theta}),
\label{eq:sllg-functional}
\end{align}
where we have defined
$\phi_{i,\mu}^{\theta}:=\phi_{i,\mu}+\theta_{i+\muhat}-\theta_{i}$.
Note that we have also introduced a normalisation factor $1/V$, where
$V$ is the lattice volume.
The stereographic gauge fixing functional becomes
\begin{align}
F_{\phi}^{s}(\theta) 
 &=-\frac{2}{V}\sum_{i,\mu}\ln(\cos(\phi_{i,\mu}^{\theta}/2)).
\label{eq:mllg-functional}
\end{align}
 A given random set of $\phi_{i,\mu}$ is called a \textit{random
orbit} or a \textit{hot configuration}. The special case when all
$\phi_{i,\mu}$ are zero is called the \textit{trivial orbit}, or
\textit{cold configuration}. The gauge-fixing conditions are, respectively,
\begin{align}
f_{i}(\theta) & =-\sum_{\mu=1}^{d}\Big(\sin\phi_{i,\mu}^{\theta}-\sin\phi_{i-\muhat,\mu}^{\theta}\Big)=0,\label{eq:any_dim_sllg_eq}\\
f_{i}^{s}(\theta) & =-\sum_{\mu=1}^{d}\Big(\tan(\phi_{i,\mu}^{\theta}/2)-\tan(\phi_{i-\muhat,\mu}^{\theta}/2)\Big)=0\label{eq:any_dim_mllg_eq}
\end{align}
 for all lattice sites $i$. The FP operator for the two gauges are
\begin{align}
\begin{split}
(M_{FP})_{i,j}=\sum_{\mu}\Big( &
 -\cos\phi_{i,\mu}^{\theta}\delta_{i+\muhat,j}
 + (\cos\phi_{i,\mu}^{\theta}+\cos\phi_{i-\muhat,\mu}^{\theta})\delta_{i,j}
 -\cos\phi_{i-\muhat,\mu}^{\theta}\delta_{i-\muhat,j}\Big)\,,
\end{split}
\label{eq:FP_op_SLLG_any_dim_in_phi_theta}\\
\begin{split}
(M_{FP}^{s})_{i,j}=\sum_{\mu}\Big( &
 -\sec^{2}\frac{\phi_{i,\mu}^{\theta}}{2}\delta_{i+\muhat,j}
 +(\sec^{2}\frac{\phi_{i,\mu}^{\theta}}{2}
  +\sec^{2}\frac{\phi_{i-\muhat,\mu}^{\theta}}{2})\delta_{i,j}
 -\sec^{2}\frac{\phi_{i-\muhat,\mu}^{\theta}}{2}\delta_{i-\muhat,j}\Big)\,.
\end{split}
\label{eq:FP_op_stereographic gauge_any_dim_in_phi_theta}
\end{align}

The boundary conditions are given by
\begin{equation}
\theta_{i+N\muhat}=(-1)^{k}\theta_{i},\quad\phi_{i+N\muhat,\mu}=(-1)^{k}\phi_{i,\mu},
\end{equation}
 where $N$ is the total number of lattice sites in the $\mu$-direction.
We have $k=0$ for periodic boundary conditions (PBC) and $k=1$ for
anti-periodic boundary conditions (APBC). With PBC there is a global
degree of freedom leading to a one-parameter family of solutions,
as all the equations are unchanged under $\theta_{i}\to\theta_{i}+\vartheta,\forall i$
where $\vartheta$ is an arbitrary constant angle. We remove this
degree of freedom by fixing one of the variables to be zero. In practice,
we set $\theta_{(N,...,N)}=0$.

From now on we concentrate only on these gauge-fixing functionals
and let the $\{\phi_{i,\mu}\}$ take random values independent of
the action. This corresponds to the strong coupling limit $\beta=0$.
Note that this is sufficient to answer the question of whether or
not $n[U]$ is orbit dependent, as every gauge orbit has a non-vanishing
weight for any finite $\beta$.

It is worth noting that the functional \eqref{eq:sllg-functional}
is identical to the hamiltonian of the random phase XY model, and
that the functional for the trivial orbit is identical to the hamiltonian
of the classical XY model in statistical physics \cite{Mehta:2009,Mehta:2010pe}.

\subsection{Analytical results in 1 dimension}

\label{sec:analytical-1d}

Here we list the available analytical results in one
dimension.\footnote{We note that some analytical results for both U(1)
  and SU(2) in 1 dimension were already found in \cite{Hetrick:1991xm}.}
\begin{enumerate}
\item For the naive functional with APBC \cite{Mehta:2009,vonSmekal:2007ns},
the minima for any orbit are $\phi_{i}^{\theta}=0$ or $\pi$ for
all $i=1,\dots,N$, the number of minima for any orbit is $2$, and
the number of stationary points is $2^{N}$.
\item For the naive functional with PBC \cite{Mehta:2009,Mehta:2010pe}
the stationary points for any orbit and any odd number of sites $N$, are given by
\begin{align}
\phi_{i}^{\theta} & =(-1)^{q_{i}}\phi_{N}^{\theta}+q_{i}\pi\,\mod2\pi,\quad & q_{i} & \in\{0,1\}\,,\label{eq:s_solutions2-1}\\
\phi_{N}^{\theta} & =\frac{N\bar{\phi}+2\pi l-\pi\sum_{i=1}^{N-1}q_{i}}{1+\sum_{i=1}^{N-1}(-1)^{q_{i}}}\,,\quad & l & \in Z\,,\label{eq:sol_1d_sol_pbc_naive}
\end{align}
 with $\bar{\phi}:=\frac{1}{N}\sum_{i=1}^{N}\phi_{i}\,$. In fact,
since we are interested in solutions of $\phi_{i}^{\theta}$ only
modulo multiples of $2\pi$, it is sufficient to consider
\begin{equation}
l\in\biggl\{1,\dots,\Bigl|1+\sum_{i=1}^{N-1}(-1)^{q_{i}}\Bigr|\biggr\}-\{0\}\,.
\end{equation}
The number of stationary points is $\sum_{j=0}^{N-1}|N-2j|\binom{N-1}{j}$,
i.e., it increases exponentially with $N$. The minima occur when
$q_{i}=0$, for all $i=1,\dots,N-1$, and $\cos\phi_{N}^{\theta}>0$.
Thus, the number of minima is bounded by $N$. The number of Gribov
copies for the compact $U(1)$ theory in one dimension thus increases
exponentially, but the number of Gribov copies in the first Gribov
region increases only linearly. In \cite{Mehta:2010pe}, it was also
shown that if $\bar{\phi}=(j+\frac{1}{2})\pi$ with $j\in\mathbb{N}$,
the FP operator is singular, i.e., for these orbits Gribov horizons
do exist. $F$ evaluated at these stationary points is
\begin{equation}
F_{\phi}(\theta)|_{\theta=\theta_{0}}=N-\cos\phi_{N}^{\theta}\sum_{k=1}^{N}(-1)^{q_{k}}.
\end{equation}
The minima of this function are when both conditions (1) $q_{k}=0$
for all $k=1,\dots,N-1$, and (2) $\cos\phi_{N}^{\theta}>0$ with
$\phi_{N}^{\theta}$ given as Eqs. \eqref{eq:s_solutions2-1} and
\eqref{eq:sol_1d_sol_pbc_naive},
are satisfied. Thus, the minima are given by
\begin{equation}
\phi_{N}^{\theta}=\frac{\sum_{i=1}^{N}\phi_{i}+2\pi l}{N}\,,\quad
\text{with}\quad
\phi_{i}^{\theta}=\phi_{N}^{\theta}\mod2\pi.
\label{eq:1D_min_naive}
\end{equation}
It follows from \eqref{eq:1D_min_naive} that function values at
the minima accumulate near zero as $N\to\infty$.  The proof of this is
given in Appendix~\ref{app:Fvals-minima}.

\item We show in Appendix~\ref{appendix:higher_dim_gen_sllg_mllg} that
for the stereographic gauge with APBC, the FP operator is generically
positive definite and hence the function $F_{\phi}^{s}(\theta)$ has
only minima. In fact, there is only one minimum for any orbit in this
case, $\phi_{i}^{\theta}=0\mod2\pi,i=1,\dots,N$.
\item For the stereographic gauge with PBC, again, the FP operator is generically
positive definite as shown in Appendix~\ref{appendix:higher_dim_gen_sllg_mllg}.
There are $N$ minima for any orbit in this case\cite{Mehta:2009,vonSmekal:pvtcommun},
\begin{align}
\phi_{i}^{\theta} & =\bar{\phi}-\frac{2\pi}{N}r\mod2\pi,\quad & r & =0,\dots,N-1,\quad & \bar{\phi} & :=\frac{1}{N}\sum_{i=1}^{N}\phi_{i}\,,
\end{align}
 i.e., the number of minima is orbit-independent and increases linearly
with $N$.
\end{enumerate}

\subsection{Some analytical results in 2 dimensions}

\label{sec:analytical}
\begin{enumerate}
\item In any dimension, for the trivial orbit, there is only one global
minimum with PBC, and two global minima with APBC, for both the naive
and stereographic gauge.

\textbf{Proof:} Consider \eqref{eq:sllg-functional} defined on
a lattice of size $N^{d}$. Let the components of $\theta$ be given
by $\{\theta_{(i_{1},\dots,i_{d})}\}$ where each $\theta_{(i_{1},\dots,i_{d})}\in(-\pi,\pi]$,
$i_{1},\dots,i_{d}$ runs from $1,\dots,N$, and $d$ is the dimension
of the lattice. It is clear that the minimum value the naive Landau
gauge-fixing functional, given in Eq.\eqref{eq:sllg-functional},
can attain is $0$. Since it is a non-negative function, when $F=0$,
the only possible configuration is when each term in $F$ is zero,
i.e.,
\begin{align}
\theta_{(i_{1},\dots,i_{d})+\muhat}-\theta_{(i_{1},\dots,i_{d})}
 & =0\quad\forall i_{1},\dots,i_{d}\in\{1,\ldots,N\},\,
 \mu\in\{1,\ldots d\},\label{eq:theta_global}
\end{align}
which is the global minima. The stereographic Landau gauge-fixing
functional is also a non-negative function and hence all the above
arguments apply. Then any global minimum has to satisfy (\ref{eq:theta_global})
and so with $\theta$ defined as above we must have 
$\theta_{(i_{1},\dots,i_{d})+\muhat}=\theta_{(i_{1},\dots,i_{d})}\mod2\pi$.
Therefore, all $\theta_{(i_1,\dots,i_d)}$ can be parameterised by only
one of them, say, $\theta_{(1,\dots,1)}=\theta_{(i_1,\dots,i_d)+\muhat}$, 
for all $i_1,\dots,i_d$ and $\mu$.

Now, for the APBC case, since $\theta_{(1,\dots,1,N+1)}=-\theta_{(1,\dots,1)}$
but also from the above arguments $\theta_{(1,\dots,1,N+1)}=\theta_{(1,\dots,1)}\mbox{mod}2\pi$.
So we have $2\theta_{(1,\dots,1)}=0\bmod2\pi$. Thus either $\theta_{(1,\dots,1)}=0$
or $\theta_{(1,\dots,1)}=\pi$. So all global minima of \eqref{eq:sllg-functional}
with APBC are lattices where all of the $\theta$-variables are zero
or all $\theta$-variables are $\pi$. With PBC, we must fix one element
on the lattice and require it to be zero. But as all elements are
equal, this means that all the elements must be zero and thus the
only global minimum of \eqref{eq:sllg-functional} with PBC is a lattice
with all $\theta$-variables set to zero.

\item For a generic orbit for compact U(1), the number of stationary points
for the naive functional is $\geq2^{N}$ for APBC and $\geq2^{N-1}$
for PBC. For the stereographic gauge, this lower bound is 1. For generic
SU($N_{c}$) with PBC, the number of stationary points is lower-bounded
by $2^{(N_{c}-1)(N-1)}$ (See Appendix \ref{app:lowerbounds} for
a proof).
\item If $\theta$ is a minimum of the naive or stereographic lattice Landau
gauge with APBC, then one can construct another minimum by adding
$\pi$ to every $\theta_{i}$.

This is easy to see in one dimension: let $\theta_{i}\to\theta_{i}+\pi$,
for $i=1,\dots,N-1$, then $\phi_{i}^{\theta}\to\phi_{i}^{\theta}$.
For $i=N$, this transformation leaves $\phi_{N}^{\theta}=\phi_{N}-\theta_{1}-\theta_{N}\to\phi_{N}-\theta_{1}-\theta_{N}-2\pi=\phi_{N}^{\theta}\bmod2\pi$.
Due to the periodicity of the trigonometric functions there is no
effect of this transformation in either the gauge-fixing equations
themselves or in the hessian. It is straightforward to extend the
same argument to any dimension.

\item For the trivial orbit in $d$ dimensions, for a symmetric $N^{d}$
lattice, if $\theta=\{\theta_{i_{1},\dots,i_{d}}\}$ is a minimum,
then $\tilde{\theta}=\pm G\theta$ is also a minimum, where $G$ is any
lattice symmetry operation, including lattice translations, rotations
and reflections (or axis permutations).
This follows straightforwardly from the symmetries of the defining
equations. Specifically, for example in two dimensions, if
$\theta=\{\theta_{ij}\}$ is a minimum of the naive or stereographic
gauge fixing functional, then $\theta^{T}=\{\theta_{ji}\}$ is also a
minimum, as the following argument shows.

Let $\theta$ be a solution of Eq.\eqref{eq:sllg-functional}, i.e.,
it satisfies the following equations for all $i,j=1,\dots,N$,
\begin{align}
\frac{\partial F}{\partial\theta_{i,j}}
 & =\sin(\theta_{i,j}-\theta_{i-1,j})-\sin(\theta_{i+1,j}-\theta_{i,j})\\
 &\;\; +\sin(\theta_{i,j}-\theta_{i,j-1})-\sin(\theta_{i,j+1}-\theta_{i,j})\notag\\
(\text{relabel } i\rightarrow j,j\rightarrow i) & =\sin(\theta_{j,i}-\theta_{j-1,i})-\sin(\theta_{j+1,i}-\theta_{j,i})\\
 & \;\;+\sin(\theta_{j,i}-\theta_{j,i-1})-\sin(\theta_{j,i+1}-\theta_{j,i})\notag\\
 & =-\frac{\partial F}{\partial\theta_{j,i}}=0\,.\notag
\end{align}

Thus, $\theta^{T}$ is a stationary point of \eqref{eq:sllg-functional}.
To prove that $\theta^{T}$ is a minimum, we note that the Hessian
\eqref{eq:FP_op_SLLG_any_dim_in_phi_theta} for the naive and similarly
for the stereographic gauge in the case of the trivial orbit are both
symmetric under the interchange of coordinates, and therefore the
hessian evaluated at $\theta$ has the same eigenvalues as the hessian
evaluated at $\theta^{T}$. Thus, if $\theta$ is a minimum, then
$\theta^{T}$ is also a minimum.

It is easy to see that this permutation symmetry holds for any
dimension, and works for both naive and stereographic gauge fixing
functionals and for stationary points in general, for the trivial
orbit.  Since there are $4N^{d}d!$ elements of the symmetry group for a $N^d$
lattice ($N^d$ lattice translations, $d!$ permutations, plus
reflection and sign change), this means that if $\theta$
is a stationary point of a $d$-dimensional functional then the function
value at $\theta$ will be $4N^{d}d!$-fold degenerate unless $\theta$ maps
onto itself under a subset of the symmetry operations.  For example,
in $d=2$ the only permutation of spatial coordinates
is a transposition, giving a degeneracy of $8N^2$, but is reduced by a
factor two for any skew-symmetric stationary point, $\theta^{T}=-\theta$.
The function value
of zero is non-degenerate, as $\theta=0$ maps onto itself under all
symmetry operations.

If we choose to impose periodic boundary conditions, then we essentially
choose the fixed site to have lattice coordinate in our case
$(N,\dots,N)$, which is fixed under any permutation and sign flip, but
not under lattice rotations or reflections.  This means that the
generic degeneracy is reduced to $2d!$.
\end{enumerate}

\section{Numerical polynomial homotopy continuation method}

\label{sec:NPHC}

In general, systems of non-linear equations are extremely difficult
to solve. However, if the non-linearity in the system is polynomial-like,
then the situation is enhanced due to the recently developed algebraic
geometry methods. In particular, we will use the so-called numerical
polynomial homotopy continuation (NPHC) method \cite{SW:95} to find
all the solutions of the gauge-fixing equations. This method was introduced
in particle physics and statistical mechanics first in \cite{Mehta:2009}
and applied in
\cite{Mehta:2009zv,Mehta:2011xs,Mehta:2011wj,Kastner:2011zz,Nerattini:2012pi,Mehta:2012qr,Maniatis:2012ex,Mehta:2012wk,Hauenstein:2012xs}.
Below,
first we show that the problem of solving the extremising equations
in terms of the $\theta$-variables can be transformed into that of
solving a system of multivariate polynomial equations. Then we describe
the numerical polynomial homotopy continuation method, which can be
used to find all the solutions of a given system of polynomial equations
numerically. Finally, we give our results for the problem at hand.

To convert the naive gauge-fixing equations, we can first use trigonometric
identities to rewrite \eqref{eq:any_dim_sllg_eq} as
\begin{equation}
\begin{split}f_{i}(c,s)=
 \sum_{\mu}\big(&c_{i}(c_{i+\muhat}\sin\phi_{i,\mu}
 \!-\!c_{i-\muhat}\sin\phi_{i-\muhat,\mu}
 \!+\!s_{i-\muhat}\cos\phi_{i-\muhat,\mu}\!+\!s_{i+\muhat}\cos\phi_{i,\mu})\\
 +&\,s_{i}(s_{i+\muhat}\sin\phi_{i,\mu}-s_{i-\muhat}\sin\phi_{i-\muhat,\mu}-c_{i+\muhat}\cos\phi_{i,\mu}-c_{i-\muhat}\cos\phi_{i-\muhat,\mu})\big),
\end{split}
\end{equation}
 where we have written $s_{i}:=\sin\theta_{i}$ and $c_{i}:=\cos\theta_{i}$.
This is merely a change of notation. However, we can now add additional
equations to the system for each site $i$, namely,
\begin{equation}
g_{i}(c,s)=s_{i}^{2}+c_{i}^{2}-1=0,\label{eq:one_dim_standard_general_poly}
\end{equation}
 The combined system of all $f_{i}(c,s)$ and $g_{i}(c,s)$ is not
just a change of notation: all the $c_{i}$ and $s_{i}$ are now algebraic
variables and the equations are multivariate polynomial equations,
i.e., the fact that $c_{i}$ and $s_{i}$ are originally $\sin\theta_{i}$
and $\cos\theta_{i}$ is taken care of by the constraint equations
(\ref{eq:one_dim_standard_general_poly}). In general, for a lattice
with $N$ lattice sites, we have in total $2N$ polynomial equations
and $2N$ variables.

To convert the gauge-fixing equations arising for the stereographic
gauge, we first simply expand \eqref{eq:any_dim_mllg_eq} using the
trigonometric identity
\begin{equation*}
\tan\frac{x+y+z}{2}
 =\frac{\sin x+\cos z\sin y+\cos y\sin z}{\cos x+\cos y\cos z-\sin y\sin z}\,,
\end{equation*}
to obtain
\begin{equation}
\begin{split}f_{i}^{s}(c,s)
 =\sum_{\mu}\big(&\frac{\sin\phi_{i,\mu}c{}_{i}
 -\cos\phi_{i,\mu}s_{i}+s_{i+\muhat}}{\sin\phi_{i,\mu}s_{i}
 +\cos\phi_{i,\mu}c_{i}+c_{i+\muhat}}
  -\frac{\sin\phi_{i-\muhat,\mu}c_{i-\muhat}-\cos\phi_{i-\muhat,\mu}s_{i-\muhat}+s_{i}}{\sin\phi_{i-\muhat,\mu}s_{i-\muhat}+\cos\phi_{i-\muhat,\mu}c_{i-\muhat}+c_{i}}\big),
\end{split}\label{eq:mllg_eq_expanded}
\end{equation}
which can again be translated into the same polynomial form as above.
Here, the difference is that the above equations are not in the `polynomial
form' due to the denominator. We can clear the denominators out by
multiplying them with the numerators appropriately and assuming that
none of the denominators are zero (such solutions can be sorted and
thrown out once all the solutions are obtained).

For example, for the trivial orbit on a one-dimensional lattice with
$N=3$ and APBC, Eq.~\eqref{eq:mllg_eq_expanded} simplifies to
\begin{align}
   \frac{s_2-s_1}{c_1+c_2}-\frac{s_1+s_3}{c_1+c_3} =
   \frac{s_3-s_2}{c_2+c_3}-\frac{s_2-s_1}{c_1+c_2} = 
   \frac{-s_1-s_3}{c_1+c_3}-\frac{s_3-s_2}{c_2+c_3} = 0\,.
\end{align}
After clearing out the denominators, the equations become
\begin{align}
\begin{split}
  -2 c_1s_1 - c_2s_1 - c_3s_1 + c_1s_2 + c_3s_2 -
     c_1s_3 - c_2s_3 &= 0\,,  \\
   c_2s_1 + c_3s_1 - c_1s_2 - 2c_2s_2 - c_3s_2 +
   c_1s_3 + c_2s_3 &= 0\,,  \\
   -c_2s_1 - c_3s_1 + c_1s_2 + c_3s_2 - c_1s_3
   -c_2s_3 - 2c_3s_3 &= 0 \,,  \\
   1 - y(c_1 + c_2)(c_1 + c_3)(c_2 + c_3) &= 0\,,
\end{split}
\end{align}
where the last equation is added to ensure that the denominators are
never zero, and $y$ is an additional variable. Thus, the final set of the
equations is in polynomial form. One can then solve this system
using the NPHC method.

\subsection{The method}

Let us consider a system of multivariate polynomial equations, say
$P(x)=0$, where $P(x)=(p_{1}(x),\dots,p_{m}(x))^{T}$ and $x=(x_{1},\dots,x_{m})^{T}$,
which is \textit{known to have isolated solutions}, e.g., the above
mentioned gauge fixing equations after eliminating the global gauge
freedom. Now, the \textit{Classical Bezout Theorem} asserts that for
a system of $m$ polynomial equations in $m$ variables, for generic
values of coefficients, the maximum number of solutions in $\mathbb{C}^{m}$
is $\prod_{i=1}^{m}d_{i}$, where $d_{i}$ is the degree of the $i$th
polynomial. This bound, the \emph{classical Bezout bound} (CBB), is
exact for generic values (i.e., roughly speaking, non-zero random
values) of coefficients, e.g., for the one-dimensional naive (or minimal)
gauge fixing equations with $N$ of lattice sites and with
APBC, this number is $2^{2N}$ (because there are $2N$ polynomials
each of which is a degree $2$ polynomial). The \textit{genericity}
is well-defined and the interested reader is referred to Ref.~\cite{SW:95,Li:2003}
for details.

Based on the CBB, a \textit{homotopy} can be constructed as
\begin{equation}
H(x,t)=\gamma(1-t)Q(x)+t\; P(x),
\end{equation}
 where $\gamma$ is a random complex number. $Q(x)=(q_{1}(x),\dots,q_{m}(x))^{T}$
is a system of polynomial equations with the following properties:
\begin{enumerate}
\item the solutions of $Q(x)=H(x,0)=0$ are known or can be easily obtained.
$Q(x)$ is called the \textit{start system} and the solutions are
called the \textit{start solutions},
\item the number of solutions of $Q(x)=H(x,0)=0$ is equal to the CBB for
$P(x)=0$,
\item the solution set of $H(x,t)=0$ for $0\le t\le1$ consists of a finite
number of smooth paths, called homotopy paths, each parameterised by
$t\in[0,1)$, and
\item every isolated solution of $H(x,1)=P(x)=0$ can be reached by some
path originating at a solution of $H(x,0)=Q(x)=0$.
\end{enumerate}
The start system $Q(x)=0$ can for example be taken to be
\begin{equation}
Q(x)=\left(\begin{array}{c}
x_{1}^{d_{1}}-1\\
\vdots\\
x_{m}^{d_{m}}-1
\end{array}\right)=0,\label{eq:Total_Degree_Homotopy}
\end{equation}
 where $d_{i}$ is the degree of the $i^{th}$ polynomial of the original
system $P(x)=0$. Eq.~(\ref{eq:Total_Degree_Homotopy}) is easy to
solve and guarantees that the total number of start solutions is $\prod_{i=1}^{m}d_{i}$,
all of which are non-singular.

One can then track all the paths corresponding to each solution of
$Q(x)=0$ from $t=0$ to $t=1$ and reach $P(x)=0=H(x,1)$. By implementing
an efficient path tracker algorithm, e.g., Euler predictor and Newton
corrector methods, all isolated solutions of a system of multivariate
polynomials system can be obtained. The complex random number $\gamma$
is crucial here: it has been shown \cite{SW:95} that for a randomly
chosen $\gamma$, there are no singularities (i.e., paths do not cross
each other) for $t\in[0,1)$. This ensures that in the end we get
all the solutions. In this respect, the NPHC method has a great advantage
over all other known methods for finding stationary points or minima.

There are several sophisticated numerical packages well-equipped with
path trackers such as Bertini\cite{BHSW06}, PHCpack~\cite{Ver:99},
PHoM~\cite{GKKTFM:04} and HOM4PS2 \cite{GLW:05,Li:2003}. They all
are available as freewares from the respective research groups. We
mainly use Bertini, HOM4PS2 and PHCpack to get the results in this
paper: for each of the systems, we use at least two of the packages
to cross-check the results.

\subsection{Results}

Before proceeding to the results, it should be noted that a solution
here means a set of values of $s_{i}$ and $c_{i}$ (for the naive
or minimal gauge) or $t_{i}$'s (for the stereographic gauge) satisfies
all of the equations with tolerances $10^{-10}$. All the solutions
come with real and imaginary parts. A solution is a real solution
if the imaginary part of each of the variables is less than or equal
to the tolerance $10^{-6}$ (below which the number of real solutions
does not change, i.e., it is robust for all the cases we consider
in this discussion). The original trigonometric equations are satisfied
with tolerance $10^{-10}$ after the $s_{i}$ and $c_{i}$ (or $t_{i}$)
are transformed back to $\theta_{i}$. All these solutions can be
further refined to an \textit{arbitrary precision}.


We present the results for the two-dimensional naive functional by
classifying the obtained solutions in terms of the number of positive
and negative eigenvalues of the corresponding Hessian matrix, or the
FP operator, because then we can use the Neuberger zero as a necessary
condition for having all the solutions. Before proceeding to the two-dimensional
case, we note that we have reproduced the known analytical results
in the one-dimensional case for both APBC and PBC naive and stereographic
gauge cases using the NPHC method, up to $N=25$ for various random
orbits.

We now explore the simplest non-trivial case in higher dimensional
lattices which is the naive functional on a $3\times3$ lattice with
the trivial orbit (TO) and 10 random orbits (ROs) (i.e., randomly
chosen $\phi_{i,\mu}\in(-\pi,\pi]$) with PBC and APBC. For these
cases, the CBB values are 65536 and 262144, respectively.

\begin{table}
\begin{center}
\begin{tabular}{|c|c|c|c|c|c|c|c|c|c|c|c|c|}
\hline
 &  &  & \multicolumn{10}{c|}{No of negative eigenvalues}\\
Orbit  & $N_{\text{tot}}$  & $N_{ns}$  & 0  & 1  & 2  & 3  & 4  & 5  & 6  & 7  & 8  & 9 \\
\hline
\hline
TO  & 3768  & 1816  & 2  & 18  & 216  & 342  & 330  & 330  & 342  & 216  & 18  & 2  \\
\hline
RO1  & 2480  & 2480  & 2  & 58  & 202  & 402  & 576  & 576  & 402  & 202  & 58  & 2 \\
\hline
RO2  & 2304  & 2304  & 10  & 36  & 148  & 382  & 576  & 576  & 382  &
 148  & 36  & 10 \\
\hline
RO3  & 2440  & 2440  & 12  & 66  & 196  & 374  & 572  & 572  & 374  & 196  & 66  & 12 \\
\hline
RO4  & 2584  & 2584  & 10  & 54  & 210  & 444  & 574  & 574  & 444  & 210  & 54  & 10 \\
\hline
RO5  & 2408  & 2408  & 2  & 44  & 202  & 404  & 552  & 552  & 404  & 202  & 44  & 2 \\
\hline
RO6  & 2672  & 2672  & 10  & 60  & 208  & 460  & 598  & 598  & 460  &
 208  & 60  & 10 \\
\hline
RO7  & 2504  & 2504  & 8  & 46  & 190  & 426  & 582  & 582  & 426  & 190  & 46  & 8 \\
\hline
RO8  & 2304  & 2304  & 6  & 48  & 152  & 362  & 584  & 584  & 362  & 152  & 48  & 6 \\
\hline
RO9  & 2352  & 2352  & 6  & 50  & 182  & 384  & 554  & 554  & 384  & 182  & 50  & 6 \\
\hline
RO10  & 2382  & 2382  & 6  & 44  & 180  & 402  & 558  & 558  & 402  &
 180  & 44  & 6 \\
\hline
\end{tabular}
\caption{Summary of the Gribov copies of the naive lattice Landau gauge for
$3\times3$ lattice, for different orbits with APBC. $N_{\text{tot}}$
is the total number of real solutions and $N_{ns}$ the number of
non-singular solutions. RO denotes random orbit while TO denotes the
trivial orbit. The number of global minima is 2 for all
orbits.}
\label{table:sllg_apbc_3x3}
\end{center}
\end{table}

\begin{table}
\begin{center}
\begin{tabular}{|c|c|c|c|c|c|c|c|c|c|c|c|}
\hline
 &  &  & \multicolumn{9}{c|}{No of negative evalues}\\
Orbit  & $N_{\text{tot}}$  & $N_{ns}$  & 0  & 1  & 2  & 3  & 4  & 5  & 6  & 7  & 8  \\
\hline
\hline
TO  & 1112  & 966  & 1  & 9  & 76  & 174  & 117  & 159  & 234  & 156  & 40 \\
\hline
TO mag  & 1034  & 1034  & 1  & 17  & 110  & 188  & 173  & 156  & 193  & 156  & 40 \\
\hline
RO1  & 480  & 480  & 3  & 15  & 50  & 95  & 121  & 116  & 65  & 14  & 1 \\
\hline
RO2  & 476  & 476  & 1  & 12  & 45  & 99  & 129  & 109  & 62  & 18  & 1 \\
\hline
RO3  & 498  & 498  & 2  & 15  & 54  & 100  & 132  & 115  & 59  & 19  & 2 \\
\hline
RO4  & 542  & 542  & 1  & 16  & 56  & 108  & 149  & 131  & 64  & 16  & 1 \\
\hline
RO5  & 470  & 470  & 1  & 9  & 46  & 105  & 138  & 110  & 49  & 11  & 1  \\
\hline
RO6  & 494  & 494  & 1  & 21  & 68  & 111  & 120  & 102  & 56  & 13  & 2 \\
\hline
RO7  & 506  & 506  & 2  & 16  & 62  & 130  & 138  & 89  & 49  & 18  & 2  \\
\hline
RO8  & 484  & 484  & 3  & 18  & 43  & 101  & 142  & 104  & 52  & 19  & 2 \\
\hline
RO9  & 484  & 484  & 2  & 9  & 49  & 114  & 145  & 104  & 44  & 15  & 2  \\
\hline
RO10  & 466  & 466  & 1  & 13  & 47  & 97  & 128  & 104  & 55  & 19  & 2 \\
\hline
\end{tabular}\caption{As Table~\ref{table:sllg_apbc_3x3}, but for PBC. TO mag field denotes
the trivial orbit with the modified functional \eqref{eq:sllg-mag-functional}
with $h=0.01$ (see main text for explanation). The number of global
minima is 1 for all orbits.}
\label{table:sllg_periodic_3x3}
\end{center}
\end{table}

The results of our runs are summarised in Tables~\ref{table:sllg_apbc_3x3}
and \ref{table:sllg_periodic_3x3} for APBC and PBC, respectively.
We find that:
\begin{enumerate}
\item For the trivial orbit with APBC, there are a total of 3768 real solutions,
i.e., Gribov copies. Out of these, there are 1952 solutions which
have zero FP determinant with tolerance $10^{-8}$. These singular
solutions lie exactly on the Gribov horizon. They can be further classified
in terms of the number of zero eigenvalues of the FP operator at each
of the solutions. This amounts to classifying the singular solutions
of the polynomial system in terms of their multiplicities using the
so-called deflation singularities technique~\cite{OjWaMi83}. The
remaining 1816 real solutions have non-singular $M_{FP}$. Similarly,
the trivial orbit with PBC has 30 singular (Gribov horizon) solutions
out of a total of 1112 Gribov copies.
\item There are no singular solutions for any of the random orbits for either
PBC or APBC. The total number of Gribov copies for each random orbit
fluctuates around $\sim2500$ for the APBC case and around $\sim500$
for the PBC case. We see that the total number of Gribov copies is
an orbit-dependent quantity in both cases.
\item All the Gribov copies (excluding the Gribov horizons) can be classified
by the number of negative eigenvalues, as shown in Tables \ref{table:sllg_apbc_3x3}
and \ref{table:sllg_periodic_3x3}. In the APBC case, one can see
a perfect symmetry among the Gribov copies in the number of solutions
classified according to the negative eigenvalues of the FP operators
evaluated at these solutions. This symmetry yields a perfect cancellation
of the signs of the FP determinants, giving rise to the Neuberger zero.
For the PBC case, there is no such manifest symmetry. However, we
can easily check that the sum of signs of the FP determinants is still
zero for all orbits, even though the total number of Gribov copies
is an orbit-dependent quantity.
\item The solutions with no negative eigenvalues are the minima. We see
that for both types of boundary conditions, the total number of (local
and global) minima is orbit-dependent. The global minima can be determined
by computing the functional $F$ with appropriate boundary conditions
for each of these minima and identifying those giving the lowest
value for $F$. We find that the number of global minima is orbit-independent.
It should be emphasised that though there are two global minima for
the APBC case, both of them are trivial copies of each other.
\item Strictly speaking, because there are singular solutions for the trivial
orbit with both APBC and PBC, Morse theory does not directly apply
here.\footnote{It is important to mention that we have observed using the
   numerical algebraic geometry methods that there is a positive
   dimensional real component of dimension at least one in the solution
   space of $3\times 3$ lattices. This is surprising because after
   eliminating the global gauge degree of freedom, there should only be
   isolated solutions left. Further studying this phenomenon and whether
   this is of any physical significance is very interesting but beyond the
   scope of the present article.}
Hence one should not expect that the sum of the signs of the
FP determinants would give the Euler characteristic in such a case,
and this can indeed be seen for the trivial orbit with PBC in Table
\ref{table:sllg_periodic_3x3}. Instead, we have to work with Morse--Bott
theory, a discussion of which is beyond the scope of this article.
Note that Schaden's equivariant construction takes this fact into
account. However, a conventional way to remove such singular solutions
is to add an external magnetic field term in the original height function,
which in our case amounts to adding an external magnetic field term
to Eq.~\eqref{eq:sllg-functional}, i.e.,
\begin{equation}
F_{\phi}(\theta)\to F_{\phi}^{m}(\theta)=F_{\phi}(\theta)+h\sum_{i=1}^{N}\cos\theta_{i}\,.\label{eq:sllg-mag-functional}
\end{equation}
 The new system of equations can again be translated to polynomial
form and solved using the NPHC method. The results for the trivial
orbit with PBC (taking $h=0.01$) are also reported in Table \ref{table:sllg_periodic_3x3},
where one can again see that the sum of the signs of the FP determinants
is zero.
\end{enumerate}

\begin{figure}[thb]
\begin{centering}
\includegraphics[clip,width=0.9\textwidth]{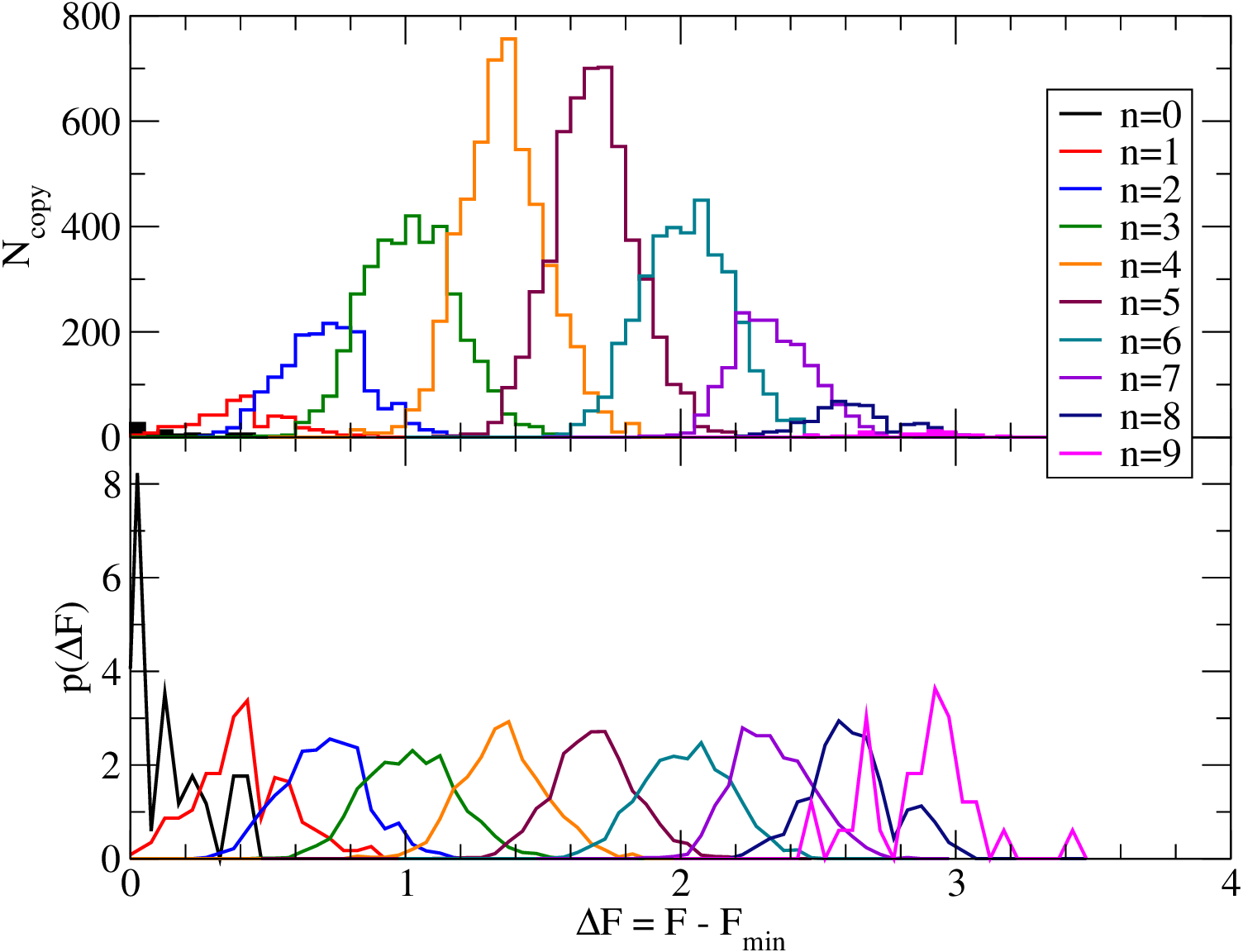}
\par\end{centering}
\caption{Distribution of the values of the gauge fixing functional
  relative to the absolute minimum for each orbit,
  $\Delta F=F-F_{\min}$, for 10 random orbits in the naive gauge on
 the $3\times3$
  lattice with antiperiodic boundary conditions, and for different
  number $n$ of negative eigenvalues.  Top: Histogram of the absolute
  number $N_{\text{copy}}$ of copies found in each bin; Bottom:
  The distribution separately normalised for each value of $n$.}
\label{fig:Fvals-3x3}
\end{figure}

From our results we can also find the distribution of functional
values $F$, broken down by the number of negative eigenvalues (ie, for
each Gribov region).  The results are shown in
Fig.~\ref{fig:Fvals-3x3} for antiperiodic boundary conditions; the
results for periodic boundary conditions are similar.  We see that the
functional values cluster around different values for different
$n$, with a peak of the distribution which is roughly linear in $n$.
However, there is a substantial overlap between the distributions for
different $n$, and in particular (because of the much larger number of
saddle points than of minima) there are many saddle points with the
same functional value as that of typical minima.

It would clearly be interesting to find out how this pattern changes
for larger lattice sizes, but this is beyond the scope of this study.

The equations for the stereographic gauge are quite difficult to
solve: the polynomial form of the equations is both dense and of high
degree.  We have only been able to find all the solutions for the
trivial orbit on the $3\times3$ lattice with periodic boundary
conditions, where the number of Gribov copies
was found to be 5256.  This is much larger than the 112 copies found
in the naive gauge, showing that the exponential suppression in 1
dimension does not carry over to higher dimensions.

\section{Numerical minimisation}

\label{sec:numerical}

So far we have collected or worked out some analytical results for
solutions in the one dimensional lattices and have used the NPHC method
to find all the solutions for the $3\times3$ lattices. For lattices
larger than $3\times3$, we have not yet been able to obtain results
using the NPHC method. To get results for the bigger lattices,
we have to rely on the traditional 'brute force' methods. Note that
for these traditional methods, there is no guarantee of finding all solutions
nor all minima of a given function. But our strategy is to first study
the cases for which we already know all the solutions: we reproduce
the results for the one dimensional lattices and for the $3\times3$
lattice using the traditional method. This gives us a hint on how
much computational effort we will require for larger lattices.

In particular, we have used the conjugate gradient algorithm~\cite{NR}
to numerically obtain minima and stationary points for both the naive
and stereographic gauge. Once the conjugate gradient algorithm has
converged we check that $\parallel\nabla_{\theta}F\parallel^{2}\leq\epsilon_{1}$
where $\epsilon_{1}$ is sufficiently small to ensure that we have
actually obtained a minimum. We reduce $\theta\bmod{2\pi}$ to make
sure the values are in the interval $(-\pi,\pi]$. We have to set
a tolerance, $\epsilon_{2}$, on how close $\theta_{i}$ has to be
to $\pi$ or $0$ to be considered either, respectively, and apply these
changes. Errors occurred when trying to determine the total number
of minima if this flag was not in place. Also, if $|\theta_{i}+\pi|\leq\epsilon_{2}$
then $\theta_{i}\to\pi$, since $\theta_{i}\in(-\pi,\pi]$, not $[-\pi,\pi]$.

\begin{figure}[htb]
 \centerline{\includegraphics[clip,width=0.8\textwidth]{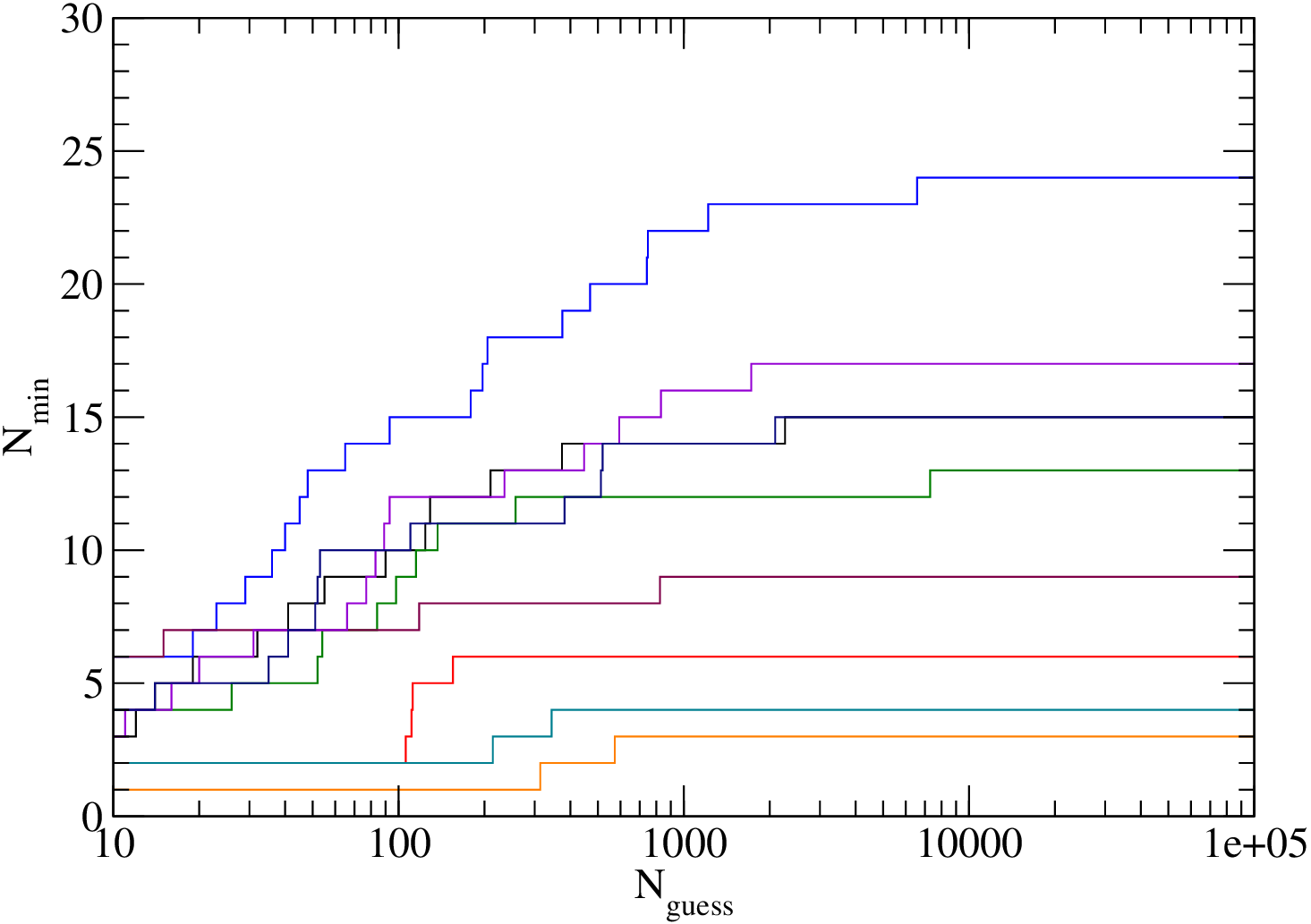}}
\caption{Number of distinct minima $N_{\text{min}}$ found as a
  function of the number of random initial guesses $N_{\text{guess}}$
  (start vectors for the CG minimisation), for 9 different random
  orbits in the naive gauge on the $5\times5$ lattice with periodic boundary
  conditions.  Note the logarithmic scale on the horizontal axis.}
\label{fig:minvsguess}
\end{figure}

To obtain {}``all'' minima, we generate a series of random initial
guesses for the variables $\theta_{i}$, and successively minimise
the given functional from each initial guess. We can then count the
number of unique minima. After increasing the number of samples sufficiently,
it is noted that in most cases the total number of unique minima that
the function converges to stops increasing and thus we can say to
have completely sampled the solution space of the function. Examples
of this are shown in Fig.~\ref{fig:minvsguess}. We note that the
number of unique minima found is usually far smaller than the number
of samples used. We call this method the Monte Carlo Conjugate Gradient
Method (MCCGM).

\subsection{1 dimension\label{sec:numerical-1d}}

\begin{figure}
\begin{center}
\includegraphics[clip,width=0.49\textwidth]{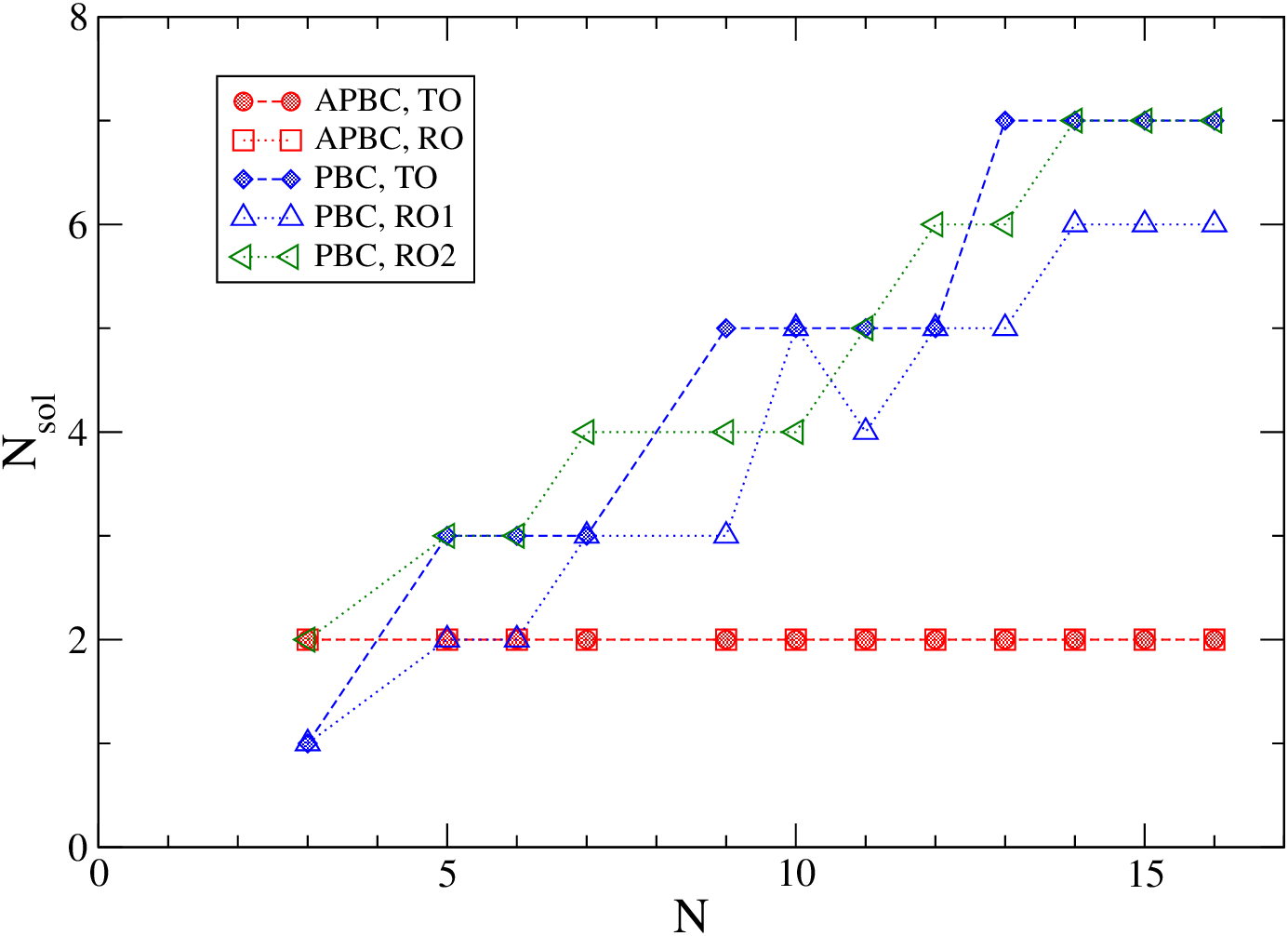}
\includegraphics[clip,width=0.49\textwidth]{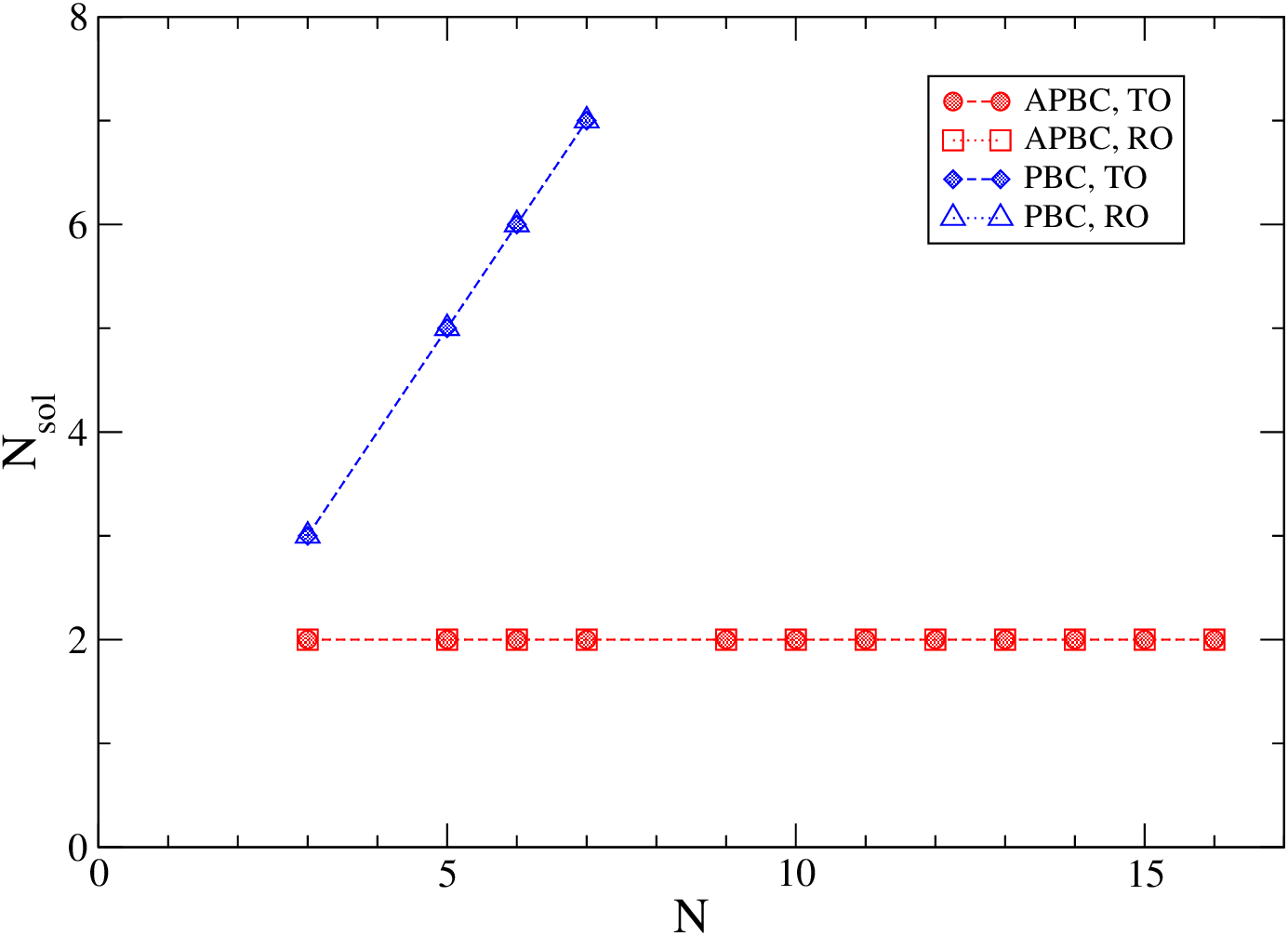}
\end{center}
\caption{The number of (local and global) minima $N_{\min}$ vs the
  number of lattice sites $N$, for naive (left) and stereographic
  gauge (right) in 1 dimension, with PBC and APBC, for the trivial
  orbit (TO) and random orbits (RO).}
\label{fig:nminima-1d}
\end{figure}

Since our method does not guarantee that we find all the minima, we
first show that we reproduce all the minima in the one-dimensional
case where rigorous analytical results are available. Indeed, we have
reproduced all the minima for lattices up to $N=1000$, for both the
naive/minimal and stereographic gauge with APBC. We also reproduced
all the minima with PBC  for up to $N=16$ for the
naive functional and up to $N=7$ for stereographic gauge. In Figure
\ref{fig:nminima-1d}, we plot the number of minima as a function
of the number of lattice sites for both gauges with both types of
boundary conditions With APBC the number of minima is always 2 for
both the stereographic and naive gauge functional. With PBC, the number
of minima is orbit-dependent, and correspondingly two possible values
are shown for each $N$. These results are in complete agreement with
the analytical results found previously.

To separate out the global minimum, we have then evaluated the
corresponding functional value at each of the local minima,
and identified the minimum (or minima) with the lowest functional
value. 
To remove any uncertainty, we have also used Matlab's in-built routines
for Simulated Annealing and Genetic Algorithm to find the global minimum
for the one-dimensional cases, and these global minima always match
the one obtained from the conjugate gradient minimisation. The global
minima from all these methods match up to $N=35$ for the naive APBC
case. For lattice sizes larger than this the Matlab minimisation routines
became inefficient both in terms of memory and computation time. The
MCCGM was found to be superior to both of these methods as it took
less time and memory to find all minima compared to the Simulated
Annealing and Genetic Algorithm routines. We find that for all the
random orbits, the number of global minima is the same (one for PBC
and two for APBC).

\begin{figure}[htb]
 \includegraphics[clip,width=\textwidth]{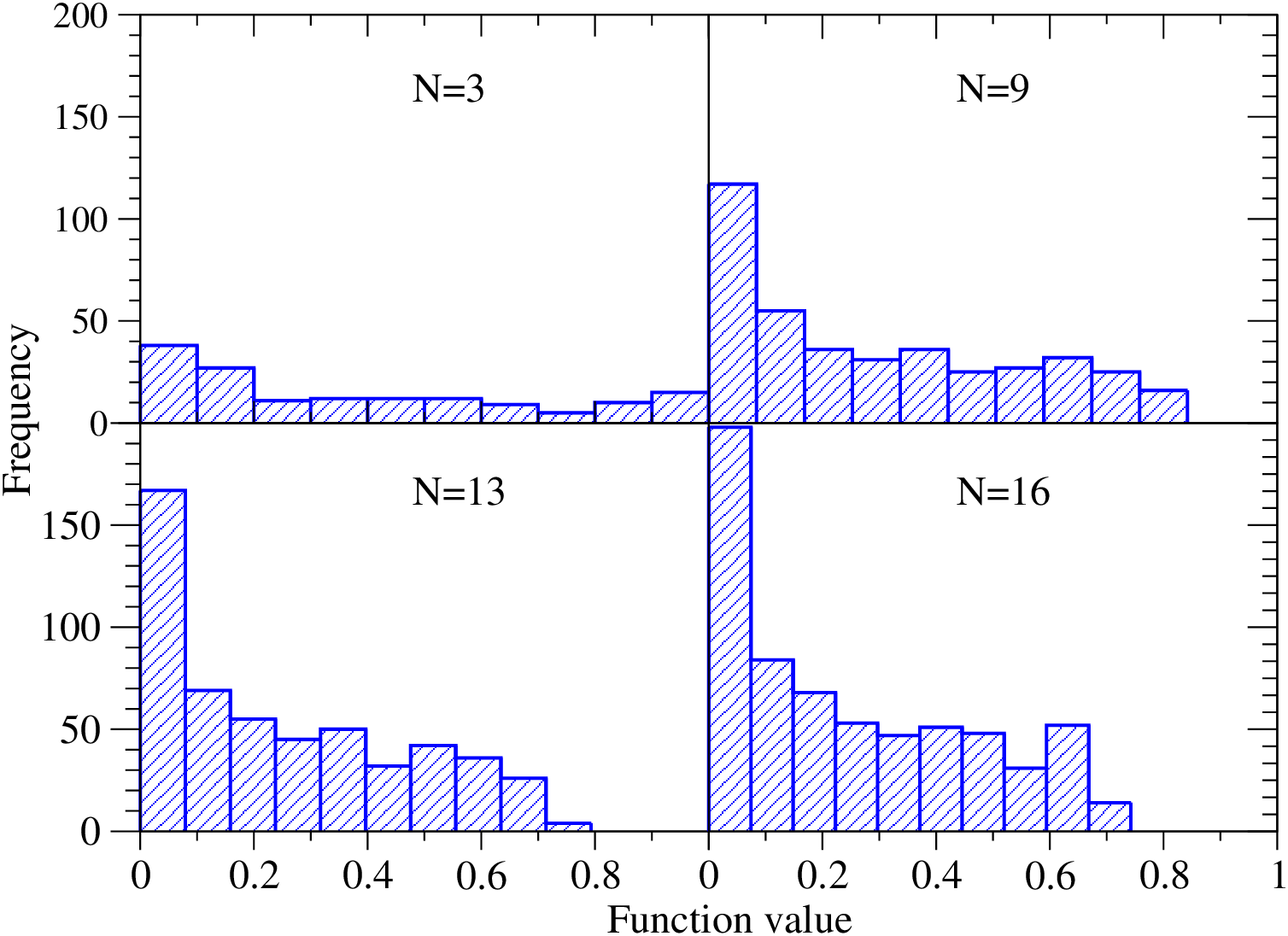}
\caption{Histogram of function values at each minimum for 100 different orbits,
for the one-dimensional naive functional with PBC.}
\label{fig:histo}
\end{figure}

In figure~\ref{fig:histo} we show the distribution of function values
at each minimum for 100 different orbits, for the one-dimensional
naive (minimal) gauge with PBC, for four different values of the lattice
size $N$. We see that for small $N$ the function values are almost
uniformly distributed between 0 and 1, but as $N$ grows the function
values tend to accumulate near 0. This is in accordance with the arguments
in Appendix~\ref{app:Fvals-minima}, that function values
accumulate near zero as $N\to\infty$.

For the stereographic
gauge, our numerical minimisation approach broke down beyond $N=16$
for PBC, in that we failed to find all the minima even when increasing
the number of initial guesses to beyond $10^{7}$ (for APBC it worked
fine up to $N=1000$). In this case each minimum is located in infinitely
deep potential wells, and which particular minimum we find is uniquely determined
by the initial guess. A strategy to select the initial guesses by
uniformly sampling these potential wells in the PBC case has proved
elusive. The same problem affects the 2-dimensional stereographic
gauge even more strongly, as we shall see in the next section.

\subsection{2 dimensions\label{sec:numerical-2d}}

\subsubsection{Naive lattice Landau gauge}
\begin{figure}[htb]
\begin{center}
\includegraphics[clip,width=0.8\textwidth]{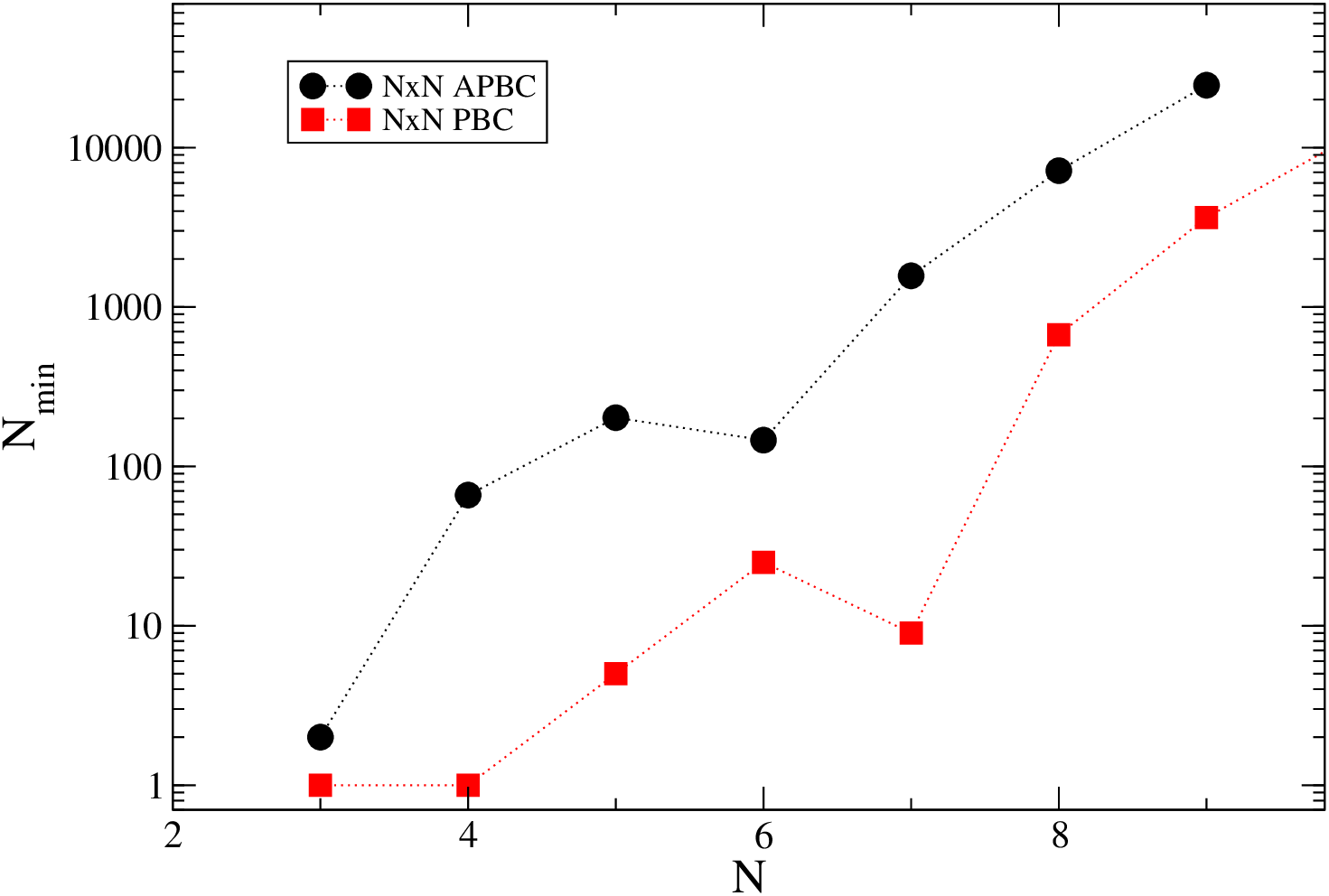}
\end{center}
\caption{Plot of the lattice size $N$ vs number of (global and local)
  minima of the naive functional in 2D, with PBC and APBC, for the
  trivial orbit. }
\label{fig:nmin-triv}
\end{figure}

\label{sec:num-sllg-2d}

Table~\ref{tab:nmin-triv} and Fig.~\ref{fig:nmin-triv} show the number
of minima $N_{\min}$ for
the trivial orbit of the naive functional with both boundary
conditions.  Because of the permutation symmetries noted in
Sec.~\ref{sec:analytical}, each minimum will in general be $n$-fold
degenerate, where $n=16N^2$
for APBC and $n=4$ for PBC.  For smaller lattice sizes in particular,
some of the minima are invariant under a subset of these permutations,
reducing the level of degeneracy.  For periodic boundary conditions,
we find a higher level of degeneracy than naively expected, suggesting
additional symmetries.  Indeed, studying the individual solutions we
find that they consist of permutations of the same numbers.  We have
been able to classify some of these permutations, but have not as yet
found any systematic pattern that is valid for generic lattice sizes.

It is clear that the number of distinct minima increases considerably
for $N>7$, but it is not possible to conclude from our results on
these small lattices whether it increases polynomially or
exponentially with $N$.

\begin{table}[ht]
\begin{centering}
\global\long\def\arraystretch{2}
\begin{tabular}{|l|rrrrrrrr|}
\hline
Lattice Size $N$  & 3  & 4  & 5  & 6  & 7  & 8  & 9 & 10 \\
\hline
$N_{\min}$ (APBC)  & 2  & 66  & 202  & 146  & 1570  & 7170  & 24626 & \\
$(N_{\min}-2)/N^{2}$  & 0  & 4  & 8  & 4  & 32  & 112  & 304 & \\
$N_{\text{distinct}}$  & 1  & 2  & 2  & 2  & 4  & 13  & 20 & \\
$N_{\min}$ (PBC)  & 1  & 1  & 5  & 25  & 9  & 671  & $\gtrsim4400$ & $\gtrsim12000$ \\
$N_{\text{distinct}}$ (PBC)  & 1  & 1  & 2  & 4  & 3  & 9  & 14
& $\geq39$ \\
\hline
\end{tabular}
\par\end{centering}
\caption{Details about the total number of minima $N_{\min}$ and distinct
(non-degenerate) minima $N_{\text{distinct}}$ for the naive functional,
trivial orbit, on $N\times N$ lattices.}
\label{tab:nmin-triv}
\end{table}

\begin{figure}[htb]
\begin{center}
\includegraphics[clip,width=0.7\textwidth]{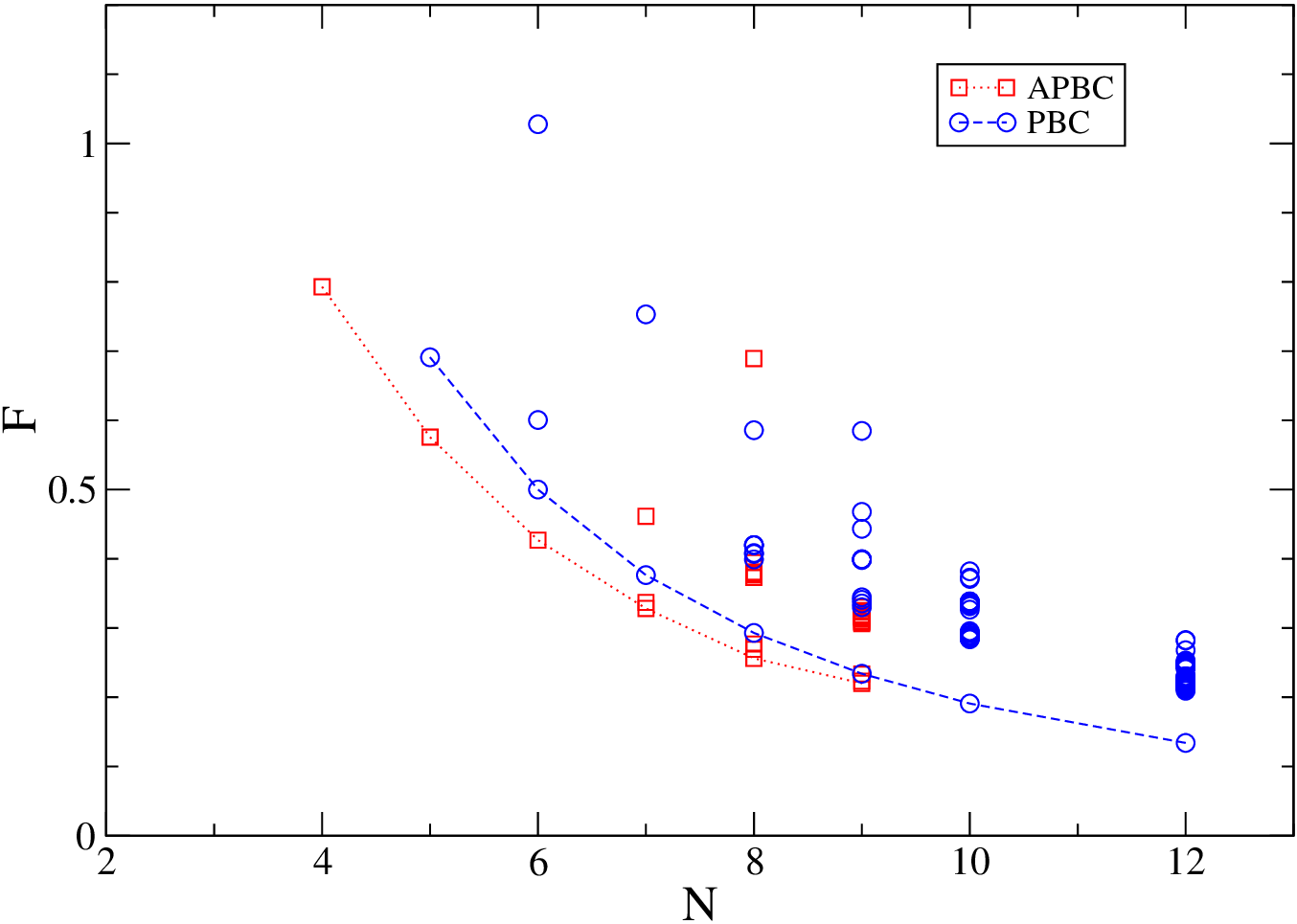}
\caption{Function
  values at the minima for the 2-dimensional naive functional,
trivial orbit, as a function of the lattice size $N$.  Note that the
  global minima at $F=0$ are not shown here.}
\label{fig:2dF}
\end{center}
\end{figure}

In Figure~\ref{fig:2dF} we plot the function values at each minimum
against $N$ for each lattice of size $N\times N$ up to a lattice of
size $12\times12$. This shows that the function values of 2-D minima
of the trivial orbit appear to decay towards zero, just as in the 1-D
case. This is again in accordance with the arguments in
Appendix~\ref{app:Fvals-minima}.  For the lowest non-trivial minimum,
which clearly can be seen to decrease with $N$, we see by inspection
that this consists of a layering of 1-dimensional minima, as described
in Appendix~\ref{app:Fvals-minima}, whose function values (at least
for the lower-lying ones) can be shown to decrease with $N$.
However, we do not have a proof that most minima approach a
function value of zero.  We shall see later that this behaviour does
not carry over to random orbits in 2 dimensions.

To study the random orbit case, we have generated $N_{\text{orb}}$
random orbits for each lattice size.  For each orbit, we have found
all the solutions, including the global minimum.  The details are
given in Table~\ref{tab:orbits-N}.

\begin{table}[htb]
\begin{centering}
\global\long\def\arraystretch{2}
\begin{tabular}{|l|lllll|}
\hline
$N$  & 3  & 4  & 5  & 6  & 7 \\\hline
$N_{\text{orb}}$  & 100  & 100  & 100  & 50  & 10 \\
$N_{\text{guess}}$ & $2\cdot10^4$ & $5\cdot10^4$ & $2\cdot10^5$
& $2\cdot10^6$ & $6\cdot10^7$ \\ \hline
\multicolumn{6}{|c|}{Periodic boundary conditions} \\\hline
$\bra N_{\text{sol}}\ket$ & 1.9 & 4.5 & 14 & 70 & 373 \\
68\% CI & 1--3 & 2--7 & 7--21 & 40--111 & 276--534 \\
$\bra F_{\min}\ket$ & 0.56\err{13}{12} & 0.48\err{8}{8} &
0.44\err{6}{5} & 0.44\err{6}{7} & 0.43\err{6}{5} \\
$\bra F_{\text{med}}-F_{\min}\ket$ & 0.12\err{12}{12}
 & 0.11\err{7}{8} & 0.15\err{7}{7} & 0.14\err{4}{4} & 0.15\err{3}{3} \\\hline
\multicolumn{6}{|c|}{Antiperiodic boundary conditions} \\\hline
$\bra N_{\text{sol}}/2\ket$ & 2.7 & 7 & 20 & 93 & 717 \\
68\% CI & 2--4 & 4--10 & 12--29 & 42--165 & 284--1493 \\
$\bra F_{\min}\ket$ & 0.43\err{10}{11} & 0.43\err{7}{8}
 & 0.41\err{6}{7} & 0.40\err{5}{4} & 0.42\err{6}{4} \\
$\bra F_{\text{med}}-F_{\min}\ket$ & 0.17\err{21}{11} & 0.15\err{9}{9}
 & 0.14\err{5}{4} & 0.14\err{4}{3} & 0.13\err{2}{3} \\\hline
\end{tabular}
\par\end{centering}
\caption{Number of random orbits $N_{\text{orb}}$ and initial guesses
$N_{\text{guess}}$ used for the different lattice sizes using the
  naive functional, together with
  results for periodic and antiperiodic boundary conditions.
  $N_{\text{sol}}$ denotes the number of minima found and 68\% CI is
  the 68\% confidence interval in the number of minima. $F_{\min}$ is
  the  absolute
  minimum of the gauge fixing functional, and its
  median value at the local minima is $F_{\text{med}}$.  The
  errors in the last digit(s) denote the 68\% confidence interval.  }
\label{tab:orbits-N}
\end{table}

\begin{figure}[htb]
\begin{center}
\includegraphics[clip,width=0.8\textwidth]{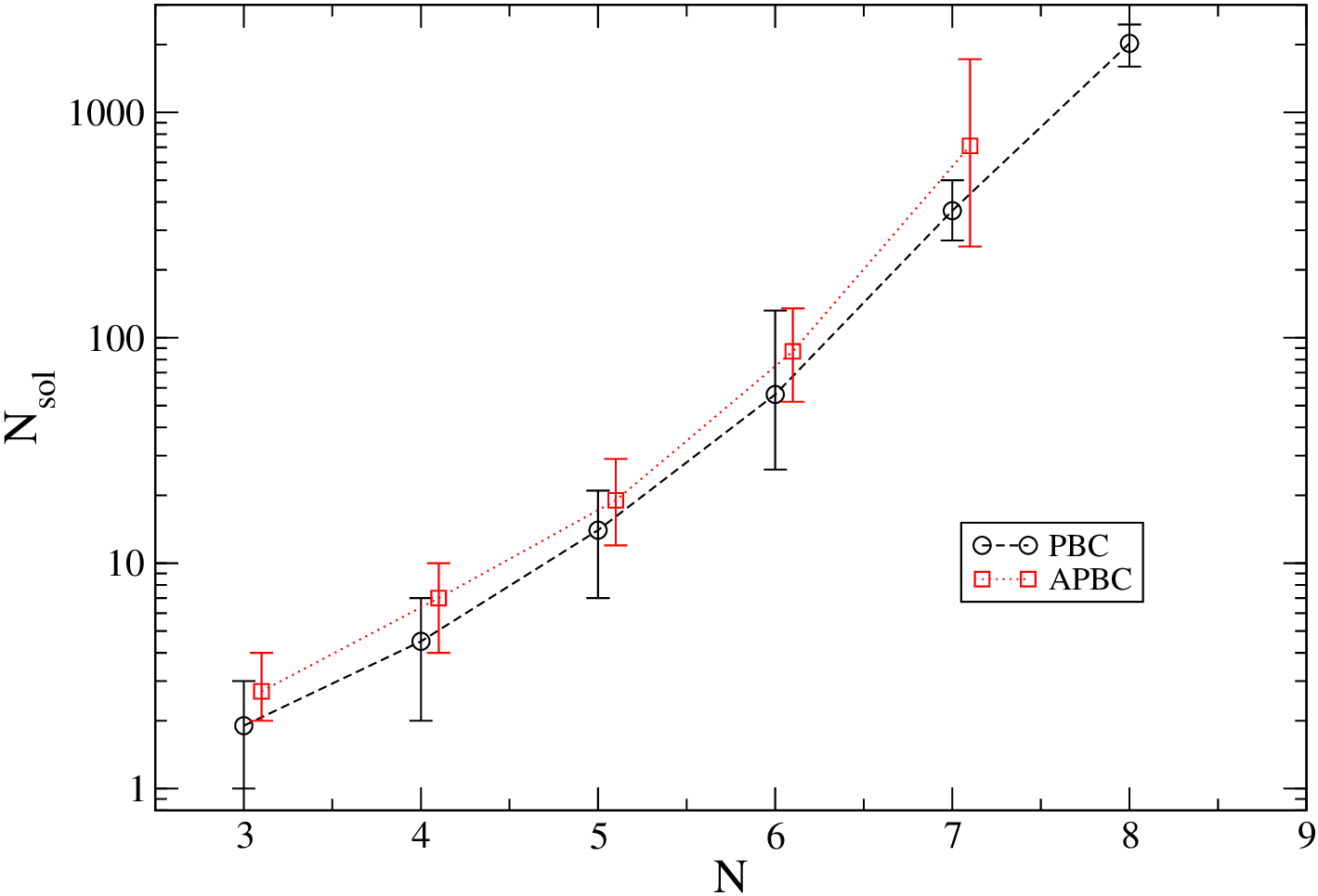}
\end{center}
\caption{Number of distinct solutions $N_{\text{sol}}$ found for
  random orbits on $N\times N$ lattices with the naive functional.  The data points denote the
  average number of solutions found for each $N$, while the error bars
denote the 68\% confidence intervals (the 16th and 84th percentiles for
the number of solutions found).  The data for antiperiodic boundary
conditions (APBC) are offset horizontally for clarity.  The $N=8$ data
  are taken from only two random orbits.}
\label{fig:nsol_rand_N}
\end{figure}

\begin{figure}[htb]
\begin{centering}
\includegraphics[clip,width=0.46\textwidth]{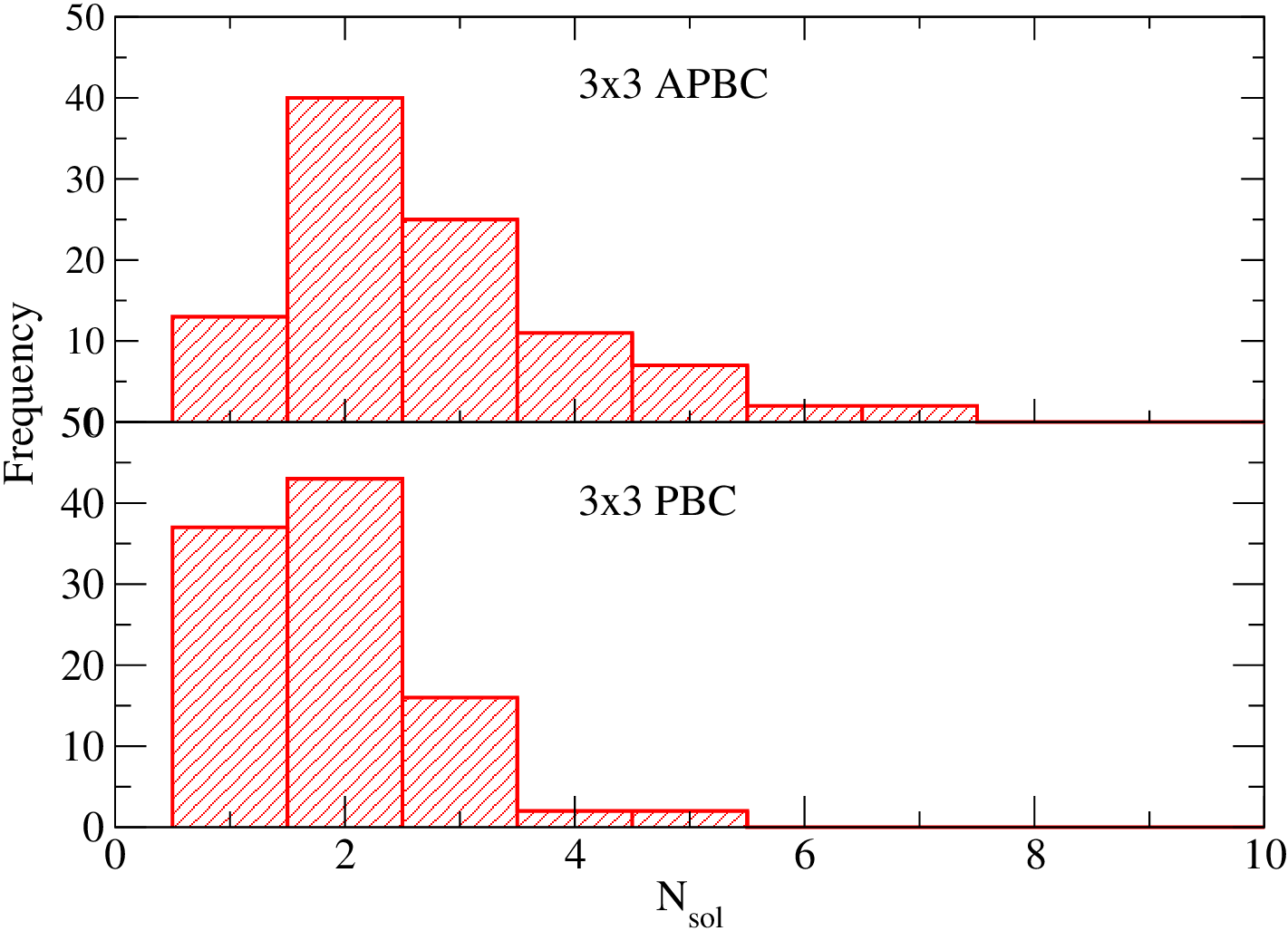}
\includegraphics[clip,width=0.46\textwidth]{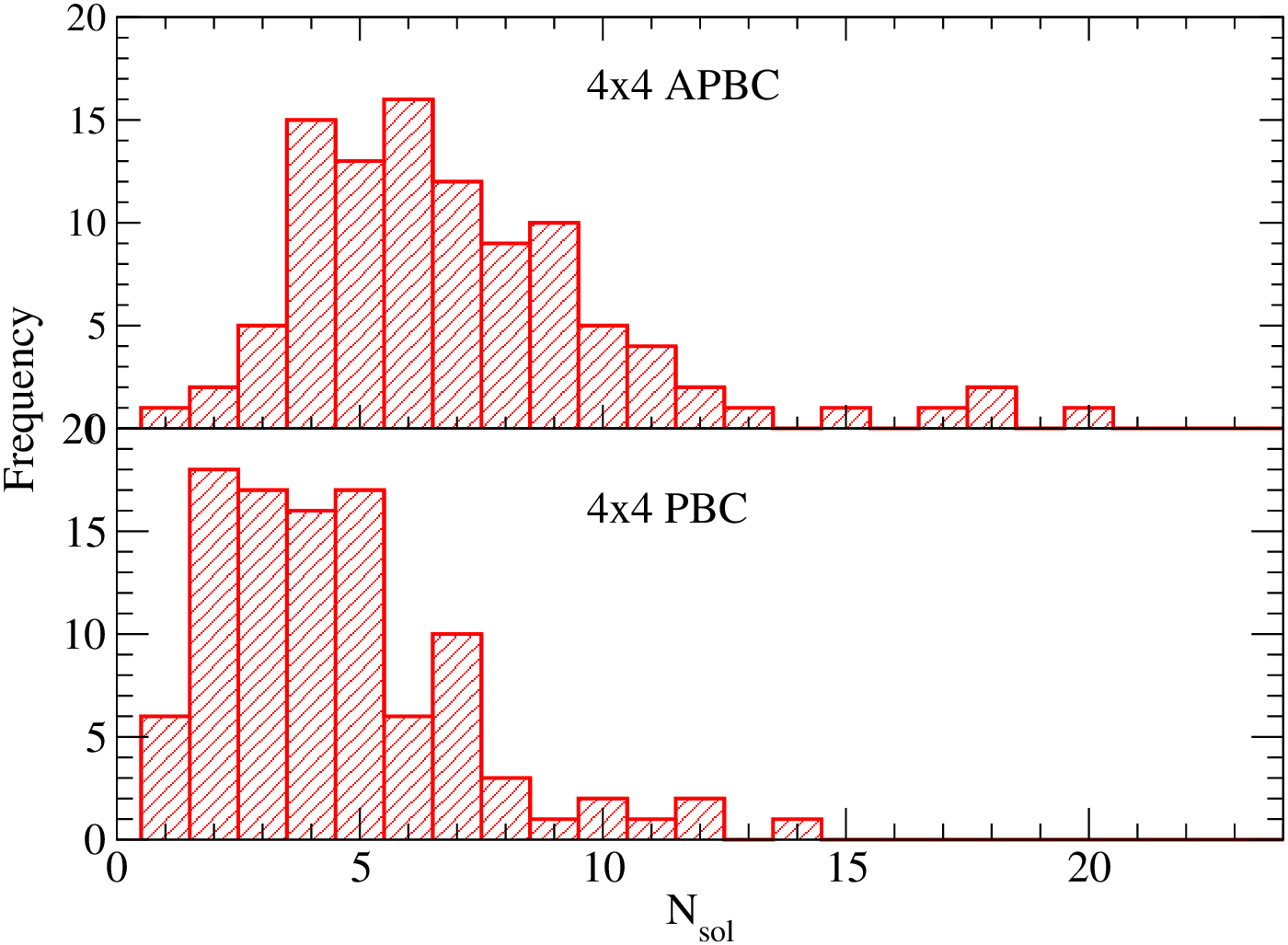} \\
\includegraphics[clip,width=0.46\textwidth]{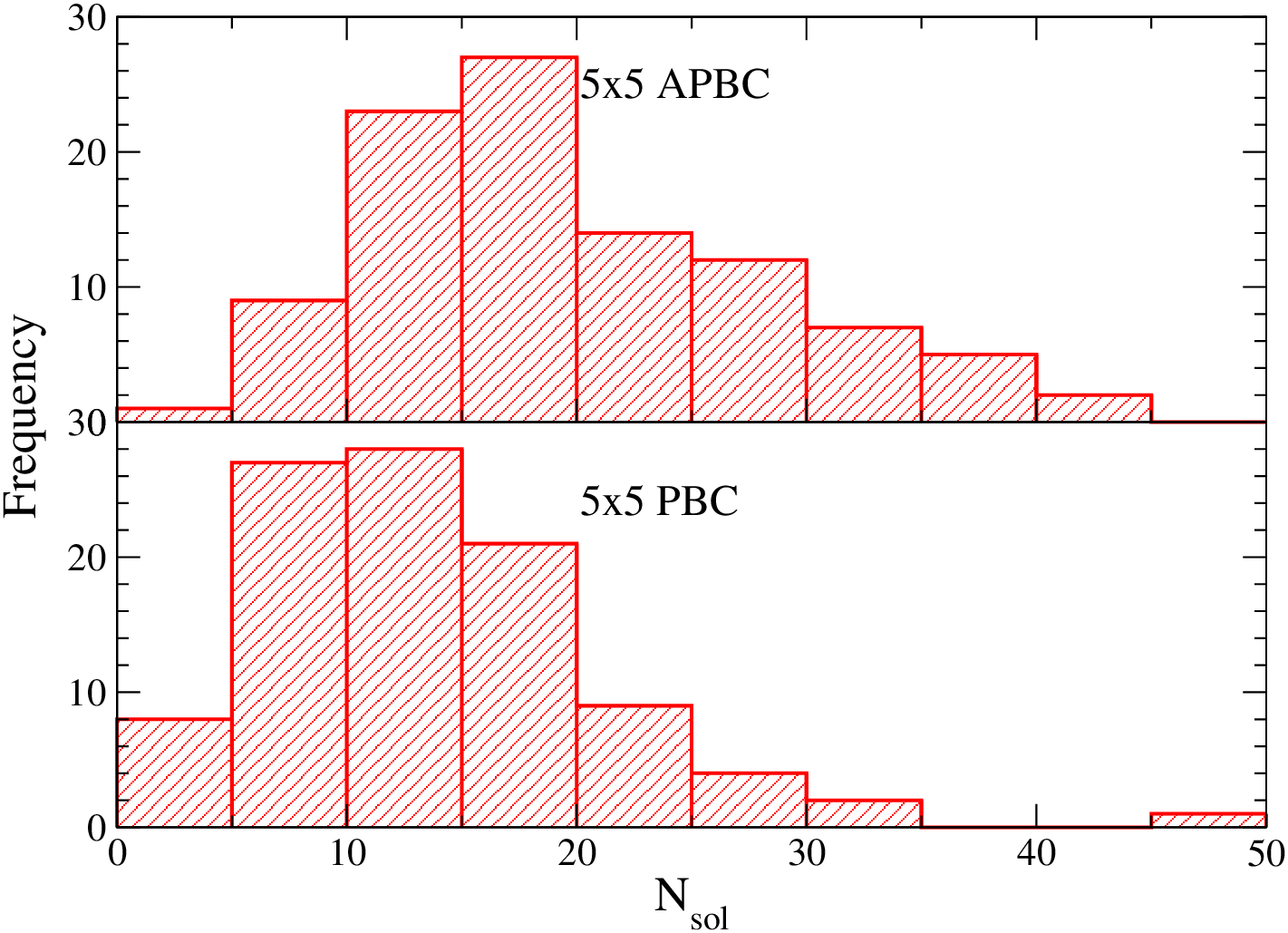}
\includegraphics[clip,width=0.46\textwidth]{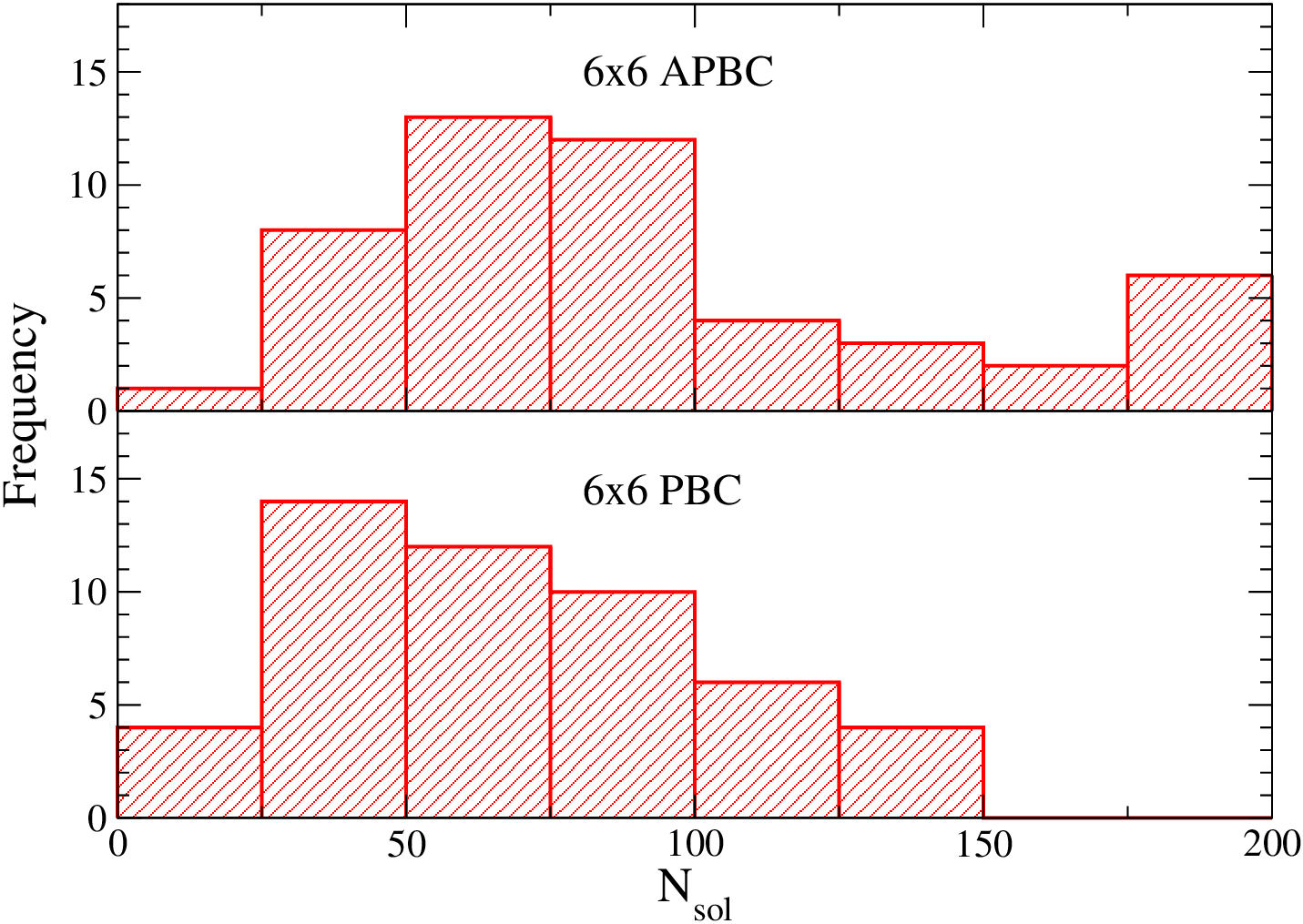}
\par\end{centering}
\caption{Distribution of the number of distinct minima
  $N_{\text{sol}}$ found for random orbits with the naive functional
  on different lattice sizes.}
\label{fig:nsol-random}
\end{figure}

Our first observation concerns the number of solutions (Gribov
copies), which, as expected, is orbit-dependent and increases with
$N$.  Unlike the trivial orbit case, all the minima are distinct for
the random orbits (or twofold degenerate for antiperiodic boundary
conditions).  The number of minima for the $3\times3$ lattice agrees
with that found using the NPHC method in section~\ref{sec:NPHC}.  Note
that because of the twofold degeneracy for APBC we have divided the
total number of solutions by 2 to factor this out.  As we can see in
Fig.~\ref{fig:nsol_rand_N} the number of solutions increases
exponentially with $N$, with roughly the same rate of increase for
periodic and antiperiodic boundary conditions.  We also found that
there is always 1 global minimum for PBC and 2 degenerate global
minima for APBC.
For a more detailed view, figure~\ref{fig:nsol-random} shows
histograms of the number of solutions for the four smallest lattice
sizes.  We do not observe any significant change in the shape of the
distribution as $N$ increases, at least for these small lattices.

We now turn to the values of the gauge fixing functional $F$ for the
different lattice sizes.  Two issues are of particular interest:
firstly, how does the absolute minimum change as a function of $N$;
and secondly, how does the spread of function values change as $N$
changes?  If the first Gribov region becomes equivalent to the
fundamental modular region in the large volume limit, as conjectured,
then we might expect the functional values to accumulate closer to the
absolute minimum for larger volumes.  We therefore determine the
absolute minimum as well as the median value of the functional at the
minimum for each random orbit.  The results are shown in
Table~\ref{tab:orbits-N}.

Unlike the case for the trivial orbit, we find no evidence that the
absolute minimum $F_{\min}$ decreases towards zero as $N$ is
increased.  Instead, $F_{\min}$ appears to converge to a value around
0.4.  This is not unsurprising, as we are working with random gauge
configurations (corresponding to the strong coupling limit), which are
not smooth and should therefore not be expected to yield smooth gauge
fixed configurations characterised by a small value for the gauge
fixing functional.

\begin{figure}
\begin{centering}
\includegraphics[clip,width=0.49\textwidth]{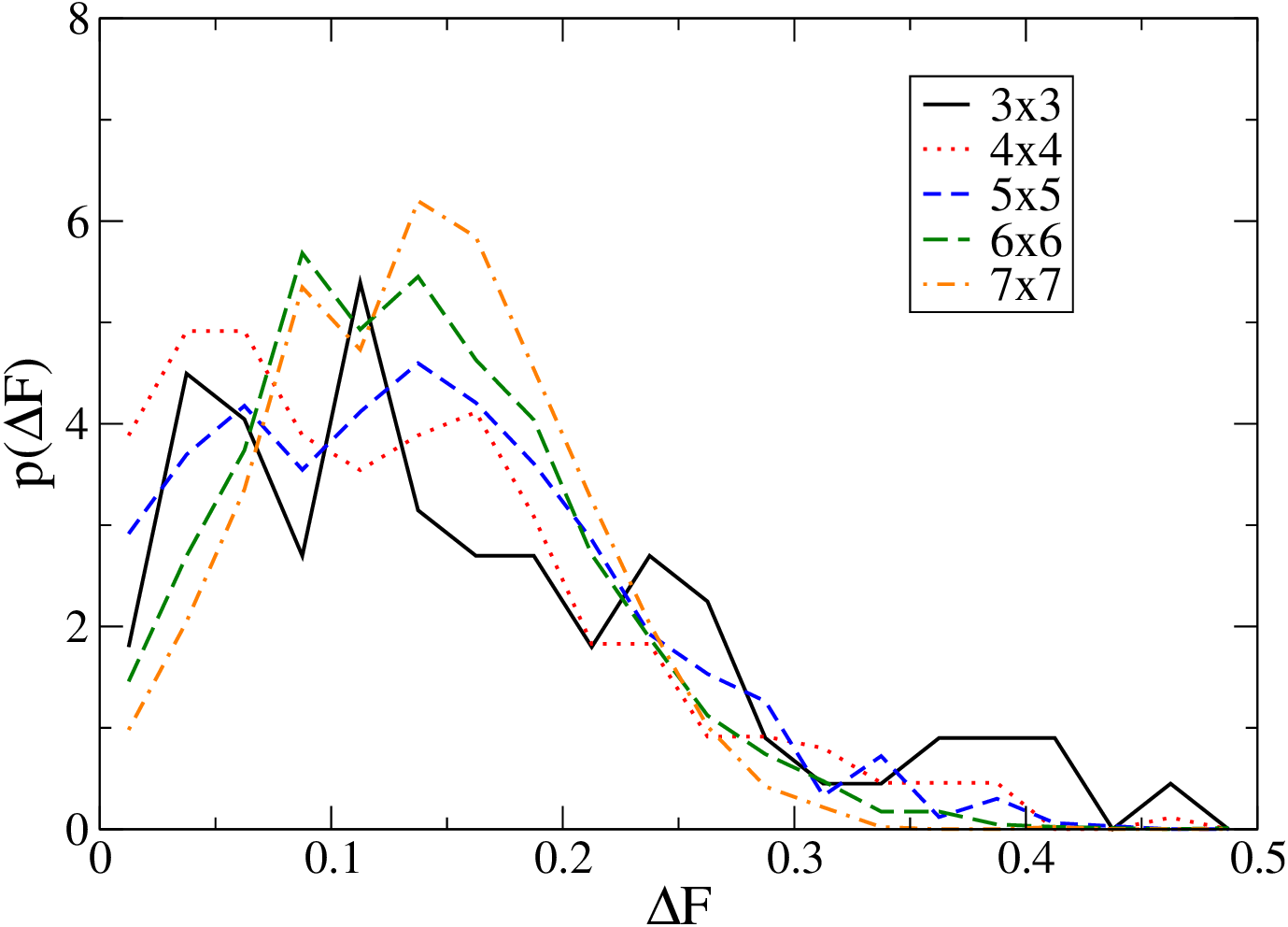}
\includegraphics[clip,width=0.49\textwidth]{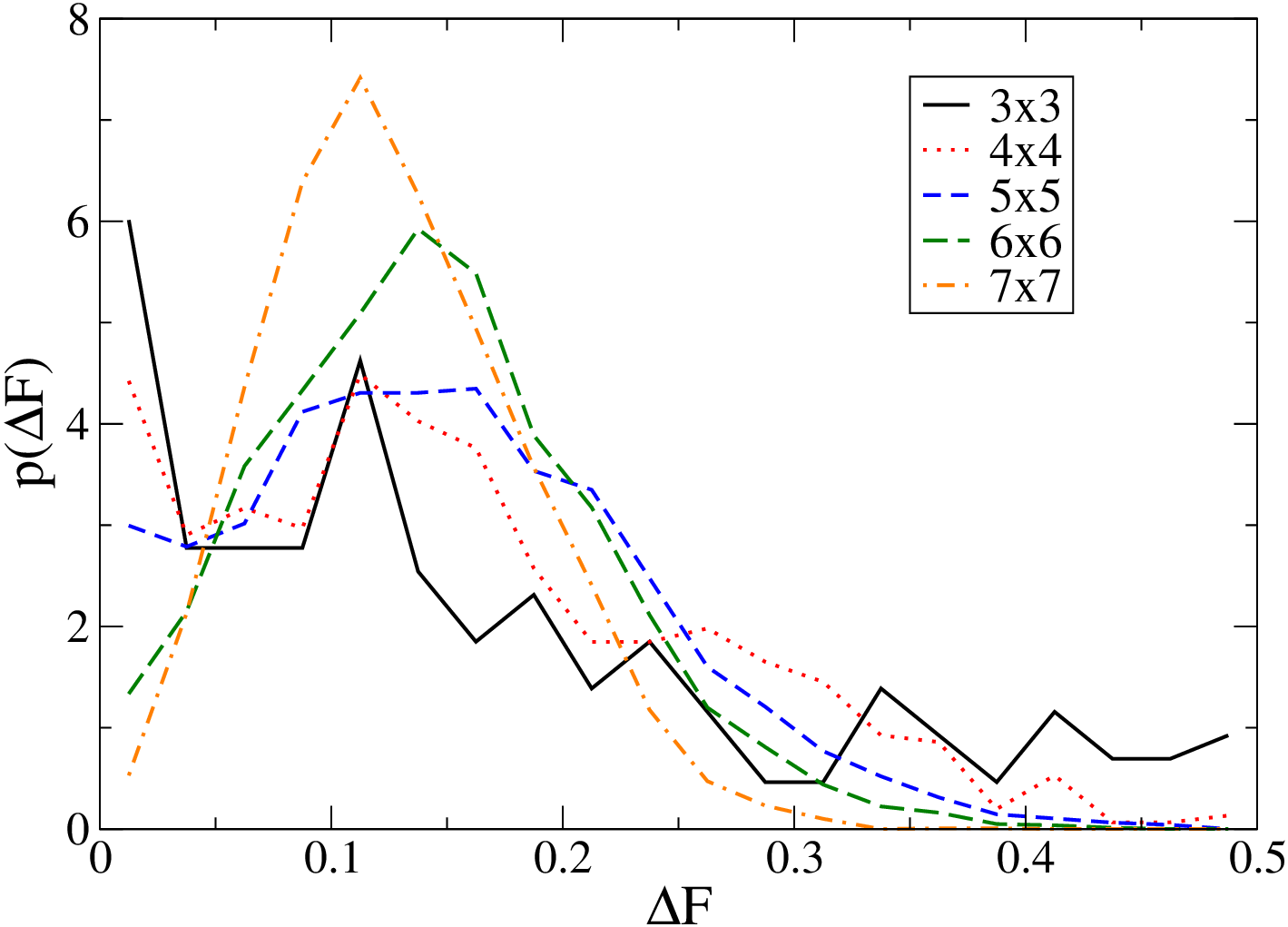}
\par\end{centering}
\caption{Distribution of the values of the naive gauge fixing functional,
  $\Delta F=F_{\text{med}}-F_{\min}$, for random orbits on different
  lattice sizes.  Left: periodic
  boundary conditions; right: antiperiodic boundary conditions.}
\label{fig:Fvals-distribution}
\end{figure}

We also find no clear evidence that the typical (or median) functional
value at the minima approaches the absolute minimum as $N$ increases.
If anything, the indication is that the opposite is the case for
periodic boundary conditions, while for antiperiodic boundary
conditions the median value may be getting closer to the absolute
minimum.  This is borne out by the distribution of functional values
relative to the absolute minimum, shown in
Fig.~\ref{fig:Fvals-distribution}.  For both boundary conditions the
distribution goes from being fairly flat to more sharply peaked as $N$
increases, but for APBC the peak appears to shift towards zero, while
for PBC the peak position stays constant or increases slightly.

As an aside, we observe that the lower minima are always the more
likely to be found by the minimisation algorithm, with the global
minimum among the first to be found.  If instead of counting each
Gribov copy equally, we weight them with how frequently they are found
by the conjugate gradient search, we find a distribution which is
peaked at zero and appears to become more strongly peaked as $N$ is
increased.  From a practical point of view this suggests that
numerical minimisation methods may have a reasonable chance of
approaching the fundamental modular region even for larger lattices,
without having to explore more than a small fraction of the Gribov
copies within the first Gribov region.

\subsection{Stereographical lattice Landau gauge}
\label{sec:num-mllg-2d}

\begin{figure}
\begin{centering}
\includegraphics[clip,width=0.8\textwidth]{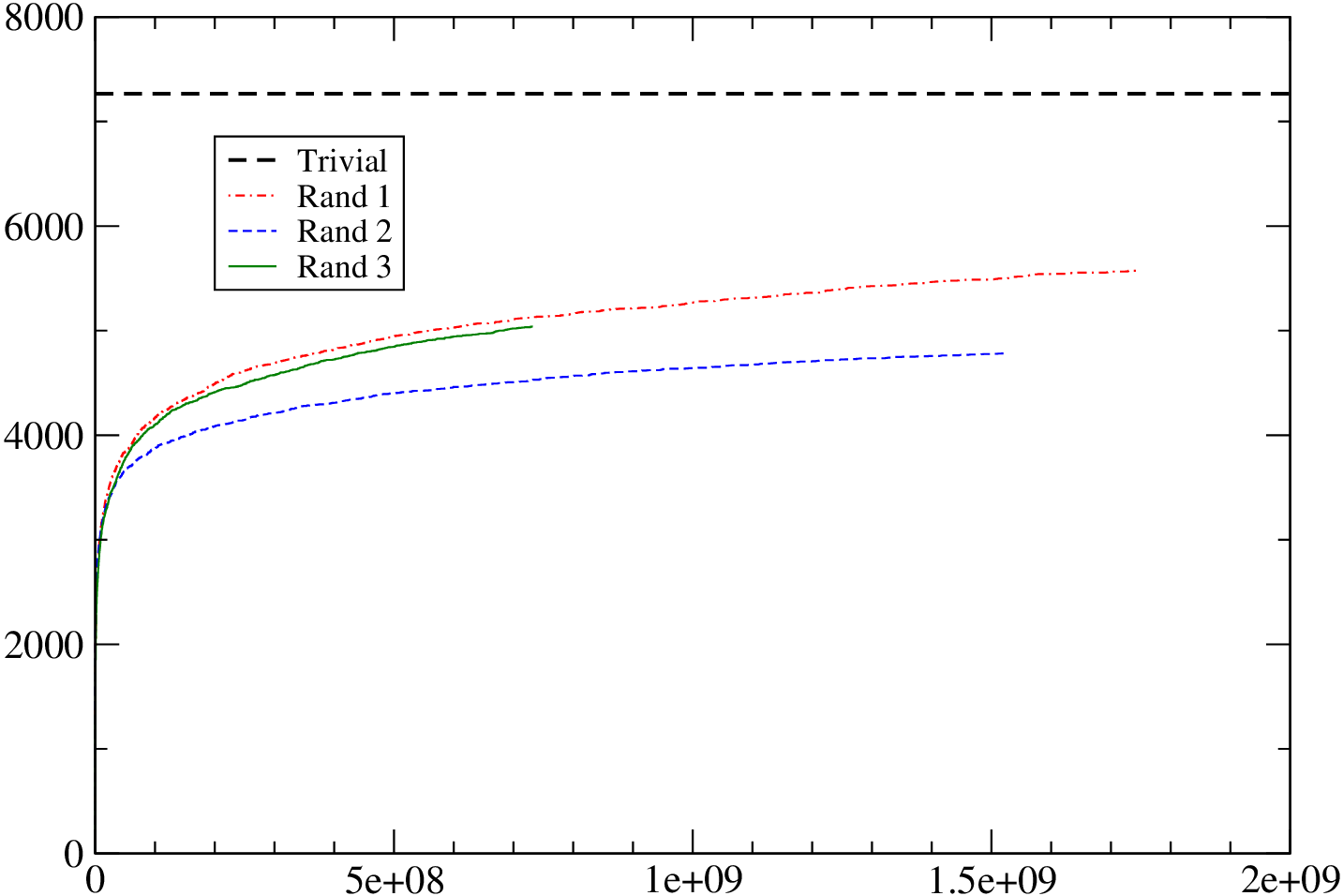}
\par\end{centering}
\caption{Number of distinct minima found for the stereographic gauge, for 3
different random orbits, as a function of the number of initial guesses
$N_{\text{guess}}$.  Also shown is the result for the trivial orbit.}
\label{fig:guessvsmin-stereo3x3}
\end{figure}

We have also attempted to use the MCCGM to find the minima of the
2-D stereographic gauge fixing functional.  In this case, we have been
restricted to a $3\times3$ lattice with APBC, and even there we have not succeeded in
finding all the minima, except for the trivial orbit.  However, we
clearly see that the number of Gribov copies (minima) for the
stereographic gauge is larger than the total number of Gribov copies
(all stationary points) for the naive lattice Landau gauge.
Moreover, the results we have obtained suggest that $n[U]$
for the stereographic gauge is orbit dependent.  As an illustration of
this, Fig.~\ref{fig:guessvsmin-stereo3x3} shows the number of distinct
minima found for three random orbits, as a function of
the number of initial guesses.
We find that the number of Gribov copies is greater than 4500 for both
random orbits considered, and 7466 for the trivial orbit. We
may compare this with the total number of Gribov copies for the naive
gauge shown in Table~\ref{table:sllg_apbc_3x3}, which ranges between
2300 and 2700 for the 10 orbits considered there.  This shows that
the exponential suppression of the number of Gribov copies observed in
1 dimension does not hold in higher dimensions, and the stereographic
gauge is thus less advantageous from this point of view than the naive
gauge.  Furthermore, the number of minima found for the two random
orbits and the trivial do
not appear to converge on a single number, although this cannot yet be
completely ruled out, as the convergence for the random orbits seems only
logarithmic in $N_{\text{guess}}$.  Similar results were obtained
using the NPHC method for 
$2\times2$ lattices with periodic boundary conditions.

\section{Discussion and conclusions}

\label{sec:discuss}

In this paper, we have studied three alternative definitions of the
lattice Landau gauge: minimal lattice Landau gauge, absolute lattice
Landau gauge, and the stereographically modified lattice Landau gauge.
We have focused on the compact $\mbox{U}(1)$ case only for this paper,
with $\beta=0$.

Our strategy is to first collect (if available) and work out the analytical
results for Gribov copies for the one dimensional lattice and, as
far as possible, for the higher dimensional lattices. For the one
dimensional case, we do have a complete understanding of Gribov copies
for the above mentioned gauges. For the two dimensional cases, we
have been able to derive some lower bounds on the number of Gribov
copies. Then, we use the recently developed numerical polynomial homotopy
continuation (NPHC) method to find all Gribov copies for a two dimensional lattice.
Here, though we could only solve small lattices of the size $3\times3$
completely, we emphasise that this is the first ever result where all
Gribov copies are found in more than one dimension. We use the NPHC
method to study the orbit-dependence of the number of Gribov copies
thoroughly for these lattices. For the bigger lattices, we move to
a traditional method, namely, Monte Carlo Conjugate Gradient Method
(MCCGM) to attempt to find `all' minima.

Our findings are summarised below.
\begin{enumerate}
\item Minimal lattice Landau gauge: The number of minima for the naive
functional in one dimension with antiperiodic boundary conditions is
orbit-independent, but in two dimensions it is orbit dependent. For
periodic boundary conditions, the number of minima is orbit-dependent in any
dimension.
\item Absolute lattice Landau gauge: The number of global minima for any
boundary conditions for the naive functional is orbit-independent,
in both one and two dimensions.  This holds for both the naive and
stereographic gauge fixing functional, in all cases where we have
obtained results.
\item Stereographic gauge: We first show that the FP operator for the stereographic
gauge is generically positive definite, and hence all the stationary
points of the corresponding gauge fixing functional are minima. We
find that the number of minima is orbit-independent in one dimension
for any boundary conditions, but appears to be orbit-dependent in two
dimensions. The number of minima for the stereographic
gauge in higher dimensions is much higher than even the number of all
the stationary points for the naive functional on the same lattice.
This is exactly opposite to the earlier claim based on the one-dimensional
results that the number of minima (Gribov copies) is exponentially
suppressed for the stereographic gauge case compared to the naive
gauge.  It is worth noting that the stereographic and naive Landau
gauge should become equivalent in the continuum limit and that the
difference between the two should be due to lattice artefacts.  This
may still pose problems for simulations at realistic couplings, however.
\item We have not found any clear evidence for the conjecture that the
  first Gribov region becomes equivalent to the fundamental modular region
in the $N\rightarrow\infty$ limit, by studying the distribution of
functional values at the minima and their median distance from the
global minimum.  It may well be that the convergence to the infinite
  volume limit is so slow that it would not be observed on these small
  lattices.  It is also possible we would need to go beyond strong
coupling and approach the continuum limit to see this.  That is beyond
the scope of this study.
\end{enumerate}

Based on our results, we conclude that it would be too cumbersome to
fix a gauge using the Faddeev--Popov 
procedure on the lattice with averaging over all Gribov copies,
due to the orbit-dependence of the Gribov copies in the minimal lattice
Landau gauge and stereographic gauge. The naive
lattice Landau gauge, as is well-known, suffers from the Neuberger
0/0 problem. Only the absolute lattice Landau gauge seems to rescue
the situation in that it evades the Neuberger 0/0 problem by construction
and the number of Gribov copies (global minima) is orbit independent.
However, finding the global minimum of such functions is in general an
NP hard problem, and it is only if typical local minima become
indistinguishable from the global minimum that this becomes a viable
alternative.  Although we find no evidence that the function values of
the minima as a whole accumulate near the global minimum, we do find
that values near the global minimum are more likely to be found.  It
remains an open question, though, whether this is still the case
for larger lattices and other gauge groups.  It is also not known whether these
near-global minima also exhibit the same qualitative behaviour for
quantities such as gauge and ghost propagators as the absolute
minimum. 

In the context of gauge fixing using a weighted average over Gribov
copies as proposed in Refs~\cite{Serreau:2012cg}, our results shown in
Fig.~\ref{fig:Fvals-distribution} suggest that the functional values
tend to a smooth distribution as $N$ grows.  This means that a
weighted average would be feasible, unlike for example the case where
there was a gap between the global minimum and all other minima,
resulting in a potentially serious overlap or ergodicity problem in
Monte Carlo simulations.  On the other hand, if the results in
Sec.~\ref{sec:NPHC} hold also for larger lattices (and other gauge
groups), then higher order stationary points are exponentially
suppressed compared to the minima, so an algorithm that samples only
minima (or only some saddle points) may provide a reasonable
approximation to the full functional integral.

We should reiterate that Gribov copies in compact U(1) gauge theory are
purely a lattice artefact with no continuum counterpart, while
SU($N_c$) gauge theories exhibit continuum Gribov copies on top of
those which are lattice artefacts.  However, understanding the nature
of Gribov copies in compact U(1) is crucial to a full understanding of
Gribov copies in SU($N_c$) theories containing compact U(1) as a
subgroup, and in particular the Neuberger 0/0 problem will be resolved
once the problem is resolved for compact U(1).  Our finding that the
stereographic gauge does not provide a satisfactory resolution of this
problem can therefore be expected to be directly applicable to other
gauge groups.  The bounds on the number of Gribov copies derived in
Appendix~\ref{app:lowerbounds} are also valid for all SU($N_c$) groups.

\section*{Acknowledgments}

This work was supported by the U.S. Department of Energy under contract
no.\ DE-FG02-85ER40237
and by Science Foundation Ireland grant 08-RFP-PHY1462.  CH
acknowledges the receipt of a SPUR scholarship.  JIS acknowledges the
hospitality of the INT at the University of Washington, where this
work was completed.
A part of this paper is based on DM's thesis and he would like to
thank Lorenz von Smekal for numerous discussions about issues related
to Gribov copies and the Neuberger 0/0 problem. DM would also like
to thank Anthony Williams for his kind support and discussions.
We would like to thank Andr\'e Sternbeck for his valuable feedback
and insights specially on the numerical issues related to Gribov copies.

\appendix

\section{Absence of Neuberger problem in stereographic gauge}

\label{appendix:higher_dim_gen_sllg_mllg}

Here we sketch the proof of the statement that there is no Neuberger
0/0 problem for the steregraphic gauge in any dimension with either
APBC or PBC. Essentially, the proof boils down to showing that the
corresponding FP operator is generically a positive definite matrix
and hence there is no cancellation among signs of the FP determinants
at Gribov copies.

We first note that the matrix $M_{FP}^{s}$ corresponds to a hessian
matrix of a functional with nearest-neighbour coupling so it can be
decomposed using projection matrices (i.e., matrices with entries in
$\{-1,0,1\}$) with only two entries per row (or column). Thus, with APBC,
the FP operator for the stereographic gauge
Eq.~(\ref{eq:FP_op_stereographic gauge_any_dim_in_phi_theta}) can be
decomposed as
\begin{equation}
M_{FP}^{s}=\sum_{\mu=1}^{d}(M_{\mu}D_{\mu}^{s}M_{\mu}^{T}),\label{eq:FP_op_any_dim_mllg}
\end{equation}
where $D_{\mu}^{s}$ are diagonal matrices with the diagonal entries
$\sec^{2}((\phi_{i,\mu}+\theta_{i+\muhat}-\theta_{i})/2)$ with $i$
running over all lattice sites and $\mu$ running over the lattice
dimension $d$. Here, the matrices $M_{\mu}$ consist of $\{-1,0,1\}$
as their entries, i.e., the $i$-th row consists of entries corresponding
to the nearest-neighbour interaction at the $i$-th site in the $\mu$-th
direction. For the APBC case, $M_{\mu}$ are non-singular matrices
because we have removed the global gauge degree of freedom using the
boundary condition. Since $M_{\mu}D_{\mu}^{s}M_{\mu}^{T}$ are positive
definite matrices for all $\mu=1,\dots,d$, according to Sylvester's
law of inertia (which states that if $A=CBC^{T}$ where $A$ is a
real symmetric matrix and $C$ and $B$ are real matrices with $B$
diagonal, then the number of positive, negative and
zero eigenvalues of $A$ is the same as those of $B$), and because
the sum of positive definite matrices is also a positive definite
matrix, $M_{FP}^{s}$ is strictly a generically positive definite
matrix. Therefore, there is no cancellation of signs of the corresponding
FP determinants, and hence, no Neuberger $0/0$ problem.

With PBC, a decomposition of the corresponding $M_{FP}^{s}$ is more
involved so we use another method here. We first separate $M_{FP}^{s}$
for different $\mu$ in Eq.~(\ref{eq:FP_op_stereographic gauge_any_dim_in_phi_theta}),
i.e., $M_{FP}^{s}=\sum_{\mu}(M_{FP}^{s})_{\mu}$ for a $d$-dimensional
square lattice. Now, we consider the quadratic form for the symmetric
matrices $(M_{FP}^{s})_{\mu}$ with an $N^{d}\times1$ vector $\vec{y}\neq\vec{0}$
whose elements are $y_{i}$ where $i=(i_{1},\dots,i_{d})$ with $i_{1},\dots,i_{d}$
running over $1,\dots,N$ for the square lattice. Thus, it is straightforward
to check that for any $\mu$,
\begin{equation}
\vec{y}^{T}(M_{FP}^{s})_{\mu}\;\vec{y}=\sum_{i}\sec^{2}\frac{\phi_{i,\mu}^{\theta}}{2}(y_{i+\muhat}-y_{i})^{2},
\end{equation}
 which is $0$ only if all $y_{i}$ are equal, which is the constant
zero mode and is strictly positive for all other cases. Also, these
$y_{i}$ are taken to be following the PBC, in this expression. Thus,
the matrix $(M_{FP}^{s})_{\mu}$ is positive semi-definite for all
$\mu=1,\dots,d$. Since the sum of positive semi-definite matrices
is a positive semi-definite matrix, $M_{FP}^{s}$ is also a positive
semi-definite matrix. Once we remove the global gauge degree of freedom
by taking one of the angles to be zero, say $\theta_{(N,\dots,N)}=0$,
then $M_{FP}$ is strictly positive definite matrix. Thus there is
no Neuberger 0/0 in the PBC case either.

\section{Lower bounds on the number of Gribov copies}

\label{app:lowerbounds}

Here, we show how the Morse theoretical interpretation of the lattice
Landau gauge can give a lower bound on the number of Gribov copies
$n$. We briefly explain the Morse indices, Betti numbers and their
relationship with $\chi(\mathbb{M})$ via the Poincar\'e polynomial
$P(z)$ of $\mathbb{M}$ along with Morse inequalities. We then obtain
an expression for the lower bound on $n$ at a given orbit and calculate%
\footnote{Here, we omit the explicit mention of the orbit-dependence of $n$.%
} it explicitly for compact U(1), SU(2) and SU($N_{c}$). The ultimate
goal of this discussion is to get a generic (orbit-independent) lower
bound on $n$ for the naive and stereographic gauge to get a guide
on our numerical results, so we ignore the explicit orbit-dependence
here.

Let $K_{i}$ be the number of critical points of a height function
$h(\vec{x})$ with its Hessian at these critical points having $i$
negative eigenvalues. Then $\chi(\mathbb{M})=\sum_{i}(-1)^{i}K_{i}$,
where now the sum over $i$ runs from $0$ to the dimension of $\mathbb{M}$.

In addition to the Euler characteristic, the Betti numbers are closely
related topological invariants of a manifold. Firstly, a homology
group is a measure of the hole structure of a manifold, or more specifically,
a topological space. There may be several homology groups of a manifold.
The $i$th Betti number $b_{i}$ is defined as the rank of the $i$th
homology group (the reader is referred to~\cite{Birmingham:1991ty}
for details on Betti numbers and their relation to the Euler characteristic).

The Poincar\'e polynomial is defined as $P(z)=\sum_{i}b_{i}z^{i}$ for
an arbitrary real variable $z$. Then, it turns out that
\begin{equation}
P(1)=\sum_{i}b_{i}\;,\,\; P(-1)=\sum_{i}(-1)^{i}b_{i}=\chi(\mathbb{M})\;,\label{eq:chi_and_b_i_general_eq}
\end{equation}
 where the sum runs from $0$ to the dimension of $\mathbb{M}$.

There are two types of inequalities, called Morse inequalities,
which relate $b_{i}$ and $K_{i}$. The weak Morse inequality states
that
\begin{equation}
\sum_{i}K_{i}\geq\sum_{i}b_{i}\label{eq:weak_Morse_inequality}
\end{equation}
 and the strong inequality states that
\begin{equation}
K_{i}\geq b_{i}\label{eq:strong_Morse_inequality}
\end{equation}
for all $i$. Due to the weak Morse inequality and using the
fact that $\sum_{i}K_{i}$ is the total number of critical points
which is $n$ for a lattice Landau gauge fixing functional, we have
\begin{equation}
P(1)=\sum_{i}b_{i}\leq\sum_{i}K_{i}=n,
\end{equation}
where for a $d$-dimensional lattice the sum runs over all the lattice
sites, i.e., $i=0,\dots,N^{d}$ for the APBC case and $i=0,\dots,N^{d}-1$
for the PBC case. Thus $P(1)$ is a lower bound on $n$. It now remains
to calculate $P(z)$ and hence ultimately $P(1)$ for the gauge-group
manifold.

To calculate the corresponding $P(z)$, first note that $P_{X\times Y}(z)=P_{X}(z)P_{Y}(z)$
for a product space manifold $X\times Y$ of $X$ and $Y$, where
$P_{X}(z)$ and $P_{Y}(z)$ are the Poincar\'e polynomials of $X$ and
$Y$ respectively. For the naive lattice Landau gauge for the compact
U(1) case with APBC, the corresponding manifold is $(S^{1})^{N^{d}}$
and the corresponding Poincar\'e polynomial is
\begin{eqnarray}
P_{(S^{1})^{N^{d}}}(z) & = & \prod_{k=1}^{N^{d}}P_{S^{1}}(z)=\prod_{k=1}^{N^{d}}(1+z)=(1+z)^{N^{d}}=\sum_{i=0}^{N^{d}}\binom{N^{d}}{i}z^{i}\nonumber \\
\therefore P_{(S^{1})^{N^{d}}}(1) & = & 2^{N^{d}}\leq\sum_{i=0}^{N^{d}}K_{i}=n\;,
\end{eqnarray}
 where we have used the fact that $P_{S^{1}}(z)=1+z$, with the Betti
numbers $b_{0}=1=b_{1}$ and all others being zero for $S^{1}$. Thus,
$n$ must be greater than or equal to $2^{N^{d}}$ in this case. Moreover,
we can verify that $P(-1)=(1-1)^{N^{d}}=0=\chi((S^{1})^{N^{d}})$
as expected. Also, we identify the Betti numbers of this $N^{d}$-torus
as $b_{i}=\binom{N^{d}}{i}$, for all $i=0,...,N$, by comparing the
above equation with Eq.~(\ref{eq:chi_and_b_i_general_eq}). Similarly,
for the PBC case, the corresponding manifold is $(S^{1})^{N^{d}-1}$,
i.e.,
\begin{eqnarray}
P_{(S^{1})^{(N^{d}-1)}}(z) & = & \prod_{k=1}^{N^{d}-1}P_{S^{1}}(z)=\prod_{k=1}^{N^{d}-1}(1+z)=\sum_{i=0}^{N^{d}-1}\binom{N^{d}-1}{i}z^{i}\nonumber \\
\therefore P_{(S^{1})^{(N^{d}-1)}}(1) & = & 2^{N^{d}-1}\leq\sum_{i=0}^{N^{d}-1}K_{i}=n\;.
\end{eqnarray}
 So, the corresponding Betti numbers for $(S^{1})^{N^{d}-1}$ are
$b_{i}=\binom{N^{d}-1}{i}$ for $i=0,...,N^{d}-1$.

For SU(2), the group manifold is $S^{3}$ and the corresponding Poincar\'e
polynomial is $P_{S^{3}}(z)=(1+z^{3})$. Thus, for the naive lattice
Landau gauge with PBC the corresponding Poincar\'e polynomial is
\begin{eqnarray}
P_{(S^{3})^{(N-1)}}(z) & = & \prod_{k=1}^{N-1}(1+z^{3})=(1+z^{3})^{N^{d}-1}\nonumber \\
\therefore P_{(S^{3})^{(N^{d}-1)}}(1) & = & 2^{N^{d}-1}\leq n^{SU(2)},
\end{eqnarray}
 giving the same lower bound for the number of Gribov copies as that
of the compact U(1) case.

For a generic SU($N_{c}$) with $N_{c}>2$, the group manifold is
$S^{3}\times S^{5}\times...S^{2N_{c}-1}$ and the corresponding Poincar\'e
polynomial is $(1+z^{3})(1+z^{5})...(1+z^{2N-1})$. Thus, for the
naive lattice Landau gauge with PBC,
\begin{eqnarray}
P_{(S^{3}\times\dots\times S^{2N_{c}-1})^{(N^{d}-1)}}(z) & = & ((1+z^{3})(1+z^{5})...(1+z^{2N_{c}-1}))^{N^{d}-1}\nonumber \\
\therefore P_{(S^{3}\times\dots\times S^{2N_{c}-1})^{(N^{d}-1)}}(1) & = & 2^{(N_{c}-1)(N^{d}-1)}\leq n^{SU(N_{c})},
\end{eqnarray}
 giving a larger lower bound than that of the compact U(1) and SU(2)
cases.

On the other hand, the corresponding manifold for the stereographic
gauge is $\mathbb{R}^{N^{d}}$ for the compact U(1) case with APBC.
Here, $b_{0}=1$, for $\mathbb{R}$, and all other $b_{i}=0$ for
$i=1,\dots,N$. So the corresponding Poincar\'e polynomial, with $P_{\mathbb{R}}(z)=1$,
is
\begin{eqnarray}
P_{\mathbb{R}^{N^{d}}}(z) & = & \prod_{k=1}^{N^{d}}(1)=1^{N^{d}}=1\nonumber \\
\therefore P_{\mathbb{R}^{N^{d}}}(-1) & = & P_{\mathbb{R}^{N^{d}}}(1)=1\leq n\;.
\end{eqnarray}
Thus, the lower bound on the number of Gribov copies is exponentially
suppressed from $2^{N^{d}}$ for the naive gauge to $1$ in the
stereographic gauge for compact U(1). Similarly, with PBC,
\begin{eqnarray}
P_{\mathbb{R}^{N^{d}-1}}(z) & = & \prod_{k=1}^{N^{d}-1}(1)=1^{N^{d}-1}=1\nonumber \\
\therefore P_{\mathbb{R}^{N^{d}-1}}(-1) & = & P_{(\mathbb{R})^{N^{d}-1}}(1)=1\leq n\;.
\end{eqnarray}
 Furthermore, the corresponding Betti numbers of $\mathbb{R}^{N-1}$
are $b_{0}=1$ and $b_{i}=0$ for $i=1,...,N$ for the APBC case and
$b_{0}=1$ and $b_{i}=0$ with $i=1,...,N^{d}-1$ for the PBC case.

\section{Gauge-fixing functional at minima}

\label{app:Fvals-minima}

Here we want to describe the function values at minima of the
gauge-fixing functional in the infinite volume limit.

In one dimension, we have the complete classification of Gribov copies
for the naive Landau gauge with PBC and odd $N$, given in
Eq.~\eqref{eq:1D_min_naive}: the minima satisfy
\begin{equation}
\phi^{\theta}_i = \phi_{N}^{\theta}
 = \bar{\phi}+\frac{2\pi l}{N} \mod2\pi \,,\quad
 l\in\{0,\dots,N-1\}\,,\quad\text{with}\quad
\bar{\phi}\equiv\frac{\sum_{i=1}^{N}\phi_{i}}{N}\notag
\end{equation}
together with the condition
$\cos\phi_{N}^{\theta}>0$.  
The functional value at the minima is then
$F_{\phi}(\theta)|_{\text{minima}}=(1-\cos\phi_{N}^{\theta})$.

By the law of large numbers, since each $\phi_i$ is independent and
identically distributed, the mean value $\bar{\phi}\to0$ as
$N\to\infty$.  Assuming that $N$ is large enough to ignore the
$\bar{\phi}$ term, we notice that the stationary points with a random
orbit in the thermodynamic limit are the same as those for the trivial
orbit. Hence in the thermodynamic limit, the first Gribov region of
every random orbit is identical to the first Gribov region of the
trivial orbit. This is a surprising result since
for finite $N$, the Gribov regions are orbit dependent.

In the limit where $\bar{\phi}\to0$, the condition
$\cos\phi^{\theta}_N>0$ becomes equivalent to $l\leq\frac{N}{4}$ or
$l\geq\frac{3N}{4}$. 
Since the function values are the same for minima in $l \in \{
0,\ldots, \frac{N}{4} \}$ and $ l \in \{ \frac{3N}{4},\ldots, N-1 \}$,
let us choose $l \in \{ 0,\ldots, \frac{N}{4} \}$.  Computing \
$F=1-\cos(2\pi l/N)$ gives us values in the interval $ [0,1] $. We have
partitioned $[0,1]$ into $\frac{N}{4}+1$ pieces (the size of the set
$\{ 0,\ldots, \frac{N}{4} \}$). Call the value of the $i$'th piece
$p_i$, where $p_i$ corresponds to the function value of $l=i$. For
sufficiently large $N$, the set of all $p_i$ will become a countably
infinite set obtaining values that appear to map the graph of $F(x) =
1-\cos(x)$ for $ 0 \le x \le \frac{\pi}{ 2}$.

However, for values of $l \ll N$, $\cos(2\pi l/N) \approx 1$ and $F
\approx 0$. In our partition of $[0,1]$ this
is equivalent to $[0,p_k] \rightarrow {0}$ as $N \rightarrow \infty$
for a particular $p_k$. As $N$ is assumed large, the rest of the
interval stretches towards zero because we divide by $N$. Hence the
function value of zero is hugely degenerate for $N$ sufficiently large
with all other function values that are not sufficiently near the
origin being unique.  We now take into account the other interval
$l\in\{\frac{3N}{4},\ldots,N-1\}$ which has the same function
values as the first interval. Therefore $F=0$ is hugely degenerate and
all other function
values not sufficiently close to the origin are $2$-fold degenerate.
Hence in the thermodynamic limit, minima occur taking function
values between $[0,1]$ but zero occurs dramatically more often.

This argument is supported by the numerical results shown in
Figure~\ref{fig:histo}. We see that even for small lattices sizes,
the function values at the minima start to accumulate near zero.

For the naive Landau gauge with PBC in two dimensions, let $\theta$ be
a stationary point. Assume that this satisfies some conditions
$c_{i,j}=\phi_{i+1,j,x}+\theta_{i+1,j}-\theta_{i,j}$ and
$d_{i,j}=\phi_{i,j+1,y}+\theta_{i,j+1}-\theta_{i,j}$. Then the
gradient equations become
\begin{equation}
 \sin(c_{i+1,j}) -\sin(c_{i,j}) + \sin(d_{i,j+1}) -\sin(d_{i,j}) = 0\,.
  \label{eq:2d_grad_equations} 
\end{equation}
Let us now assume that $c_{i,j}, d_{i,j}$ satisfy the one dimensional
gradient equations
\begin{align}
 \sin (d_{i,j+1})  -\sin (d_{i,j}) & =0 \nonumber\,, \\
 \sin (c_{i+1,j})  -\sin (c_{i,j}) & =0\,. \label{eq:1d_grad}
\end{align}
Then these satisfy the $2$-dimensional gradient equations, and we have
found stationary points for which the gradient
equations decouple into lower dimensional ones. For a particular
orbit, we may then find a higher dimensional stationary point by solving the
one dimensional gradient equations for each hypersurface with $j$
fixed in $c_{i,j}$ and $i$ fixed in $d_{i,j}$. However, optimising the
one dimensional gradient equations for $c_{i,j}$ with $j$ fixed fixes
the $j$'th column of lattice sites. With $j \in \{ 1, \ldots, N\}$,
this fixes all lattice sites. Therefore, we cannot vary the
$\theta_{i,j}$ in order to optimise $d_{i,j}$ for fixed $i$ with a
random orbit. We must impose the trivial orbit. The gradient equations
are solved only if $d_{i,j+1} = c_{G ( i+1,j )}$ where $G$ is some
lattice symmetry which imposes $\sin(d_{i,j+1})=\sin(d_{i,j})$ so that
the gradient equations are simultaneously optimised. Now we can
optimise each hypersurface $c_{i,j}$ for fixed $j$ but it is only the
subset of these SPs which simultaneously imposes the gradient
equations of $d_{i,j}$ that satisfy the two dimensional gradient
equations. Using the analytic formula for one dimensional SPs with
the trivial orbit from section ~\ref{sec:analytical-1d}, we can find
these two dimensional SPs. As noted, the minima occur at all $q_k =0$
and thus the two dimensional minima we can construct are
\begin{align}
c_{i,j} =\frac{2 \pi l}{N} \,,\quad
d_{i,j} = c_{G (i,j)} =  \frac{ 2 \pi G(l)}{N}\,,\quad
\end{align}
where $l\in \{ 0,\ldots, \frac{N}{4},\frac{3N}{4},\ldots, N-1 \}$.

We can explicitly construct some of these minima. With the above
conventions, first note that requiring $d_{i,j}=0$ ($G(l)=0$)
simultaneously solves the gradient equations for each $l$. This is
easily visualised by first fixing a lattice site to impose boundary
conditions and then layering a fixed hypersurface of a $1$-dimensional
SP in a constant direction until a $2$-dimensional lattice is
filled. Putting this back into the definition of $F$ it is easy to see
that, in this specific case, the two dimensional minima has the same
function value as the one dimensional minima.  The Hessian (FP
operator) in this $G(l)=0$ case is given by 
\begin{equation}
M_{FP}
 = \frac{\partial^2{F}}{\partial{\theta_{i,j}}\partial{\theta_{k,l}}}
 = (M_{FP}^{1D})_{i,k}\delta_{j,l}
  - ( \delta_{j+1,l} + \delta_{j-1,l} - 2\delta_{j,l})\delta_{i,k}
 \label{eq:2d_hessian} 
\end{equation}
where $M_{FP}^{1D}$ is the Hessian of the one dimensional minima,
which is a $N \times N$ matrix. From~\eqref{eq:2d_hessian} we see that
$M_{FP}^{1D}+2 I$ occur in blocks along the main diagonal. The other
non zero terms come from the second term in~\eqref{eq:2d_hessian}. They
are upper and lower diagonal $-1$ terms starting at the $(1,N+1)$ and
$(N+1,1)$ matrix element. Boundary conditions give diagonal $-1$ terms
starting at the $(1,N^2 -N )$ and $(N^2 -N,1)$ matrix element. To
fully impose PBC we must remove one linearly dependent gradient
equation. This reduces the Hessian to a $(N^2-1) \times (N^2-1) $
matrix. These minima were observed numerically. More minima can be
constructed by defining $G$ and we can find others by the permutation
symmetry as described in 
section~\ref{sec:analytical}. Therefore we can construct multiple sets
of two dimensional minima were each set is in a one-to-one
correspondence with the set of all one dimensional minima.

Since we can construct a number of $2$-dimensional minima from each
one dimensional minimum, the two dimensional first Gribov region will
contain multiple analogues of the one dimensional first Gribov
region. By the above arguments, since most function values of the
$1$-dimensional minima accumulate at zero in the thermodynamic limit,
this behaviour must be imitated locally by the $2$-dimensional first
Gribov region. As this construction of minima can be extended to
higher dimensions, this argument generalises.

We cannot comment on the global behaviour of the two dimensional first
Gribov region as an analytic classification of all two dimensional
minima remains elusive. However, the numerical results from
Figure~\ref{fig:2dF} show that the function values of $2$-dimensional
minima with a trivial orbit appear to decay towards zero, similar to
the one dimensional case. This suggests that the imitation may be
global for the trivial orbit. Since this construction only works for
the trivial orbit, there is no reason to suggest that the two
dimensional first Gribov region for a random orbit behaves locally
like the one dimensional first Gribov region. Indeed, this difference
between the trivial and random orbit case was observed in
section~\ref{sec:num-sllg-2d} and particularly in
Table~\ref{tab:orbits-N}.

\bibliographystyle{JHEP-2}
\bibliography{bibliography}
\end{document}